\documentclass[aps, prb, showpacs, twocolumn, amsmath, letterpaper]{revtex4-2}

\usepackage{graphics}
\usepackage{graphicx}

\usepackage{amssymb}
\usepackage{bm}

\usepackage[english]{babel}

\usepackage{amsmath}
\usepackage{graphicx}
\usepackage{wrapfig}
\usepackage[colorlinks=true, allcolors=blue]{hyperref}

\begin{document}

\title{The effect of the interlayer ordering on the Fermi surface of Kagome superconductor CsV$_3$Sb$_5$ revealed by quantum oscillations}

\author{Christopher Broyles$^{1}$, Dave Graf$^{2}$, Haitao Yang$^{3,4,5,6}$, Hongjun Gao$^{3,4,5,6}$, Sheng Ran$^{1}$}
\affiliation{$^1$ Department of Physics, Washington University in St. Louis, St. Louis, MO 63130, USA
\\$^2$ National High Magnetic Field Laboratory, Florida State University, Tallahassee, FL 32313, USA
\\$^3$ Beijing National Center for Condensed Matter Physics and Institute of Physics, Chinese Academy of Sciences, Beijing 100190, PR China
\\$^4$ School of Physical Sciences, University of Chinese Academy of Sciences, Beijing 100190, PR China
\\$^5$ CAS Center for Excellence in Topological Quantum Computation, University of Chinese Academy of Sciences, Beijing 100190, PR China
\\$^6$ Songshan Lake Materials Laboratory, Dongguan, Guangdong 523808, PR China
}

\begin{abstract}
The connection between unconventional superconductivity and charge density waves (CDW) has intrigued the condensed matter community and found much interest in the recently discovered superconducting Kagome family of AV$_3$Sb$_5$ (A = K, Cs, Rb). Xray diffraction and Raman spectroscopy measurements established that the CDW order in CsV$_3$Sb$_5$ comprises of a 2x2x4 structure with stacking of layers of star-of-David (SD) and inverse-star-of-David (ISD) pattern along the $c$-axis direction. Such interlayer ordering will induce a vast normalization of the electronic ground state; however, it has not been observed in Fermi surface measurements. Here we report quantum oscillations of CsV$_3$Sb$_5$ using tunnel diode oscillator frequency measurements. We observed a large number of frequencies, many of which were not reported. The number of frequencies can not be explained by DFT calculations when only SD or ISD distortion is considered. Instead, our results are consistent with calculations when interlayer ordering is taken into account, providing strong evidence that CDW phase of CsV$_3$Sb$_5$ has complicated structure distortion which in turn has dramatic effects on the Fermi surface properties. 

\end{abstract}

\maketitle{}

The Kagome lattice is an ideal environment to host topological physics within the strong electron correlation regime due to its special geometric structure, promoting a destructive quantum phase interference of electron hopping paths~\cite{Lin_1994,https://doi.org/10.48550/arxiv.1708.04448}. Tight binding model predicts flat bands together with Dirac cones, that are protected by the lattice symmetry~\cite{PhysRevB.80.113102}. The recent discovery of superconductivity in the quasi-2D Kagome metal family of AV$_3$Sb$_5$ (A = K, Cs, Rb) has introduced a new material set to study superconductivity and topology in the strong correlation regime~\cite{PhysRevMaterials.3.094407}. Theoretical calculations and angle-resolved photoemission spectroscopy (ARPES) have shown that these compounds host topological Dirac surface states with a $Z_2$ topological invariant near the Fermi energy~\cite{PhysRevMaterials.3.094407,Tan_2021,PhysRevLett.125.247002,PhysRevX.11.031050,PhysRevB.104.L161112,PhysRevX.11.031050,PhysRevLett.127.236401,PhysRevB.103.L241117}. A robust zero-bias conductance peak has been observed inside the superconducting vortex in CsV$_3$Sb$_5$, making it an ideal candidate for topological superconductor~\cite{Liang_2021}. Furthermore, signatures of spin-triplet pairing and an edge supercurrent in Nb/K$_{1-x}$V$_3$Sb$_5$ devices have been reported~\cite{https://doi.org/10.48550/arxiv.2012.05898}.

In the normal state, this family of materials exhibit clear transport and magnetic anomalies at temperatures between 78 – 104~K due to the formation of charge density wave (CDW) order~\cite{PhysRevMaterials.3.094407,PhysRevLett.125.247002,https://doi.org/10.48550/arxiv.2104.05556,Qian_2021}. Under applied pressure, the transition temperature of the CDW phase, $T_{CDW}$, decreases while the superconducting transition temperature, $T_c$, shows an unusual multiple-dome feature with a significant enhancement to about 8~K at 2~GPa, indicating a competition between CDW and superconductivity~\cite{Yu_2021,PhysRevLett.126.247001,Wang_2021}. Growing evidence has been shown that the CDW phase is highly unconventional with a chiral charge order, giving rise to the observed anomalous Hall effect in the absence of long-range magnetic order~\cite{osti_1821432,PhysRevB.104.075148,PhysRevB.104.L041103}. Xray diffraction and scanning tunneling microscopy (STM) measurements have revealed the formation of a 2 × 2 superlattice within the plane in the CDW phase~\cite{hu2022coexistence,li2022emergence,PhysRevB.104.075148,PhysRevX.11.031026,jiang2021unconventional}. In addition, another 4$a_0$ unidirectional superlattice has also been observed at low temperatures~\cite{PhysRevB.104.075148,hu2022coexistence,10.1088/1674-1056/ac5888, Chen_2021}. 

Despite the extensive studies of the CDW phase, its nature remains elusive. Theoretical calculations find that the CDW transition is related to breathing-phonon modes of the Kagome lattice and electronically mediated by the Fermi surface instability~\cite{PhysRevLett.127.046401}. Phonon band structure of AV$_3$Sb$_5$ exhibits softening acoustic phonon modes at the Brillouin zone boundary near $M$ and $L$ points, indicating strong structural instability~\cite{PhysRevB.105.L140501}. $M$-point soft mode corresponds to a breathing phonon of V atoms in the Kagome lattice~\cite{PhysRevB.105.L140501}. Breathing in and out lead to two different structures, the Star of David (SD) structure where V atoms move away from the center and the inverse Star of David structure (ISD) with an inverse deformation~\cite{PhysRevMaterials.5.L111801}. Both SD and ISD exhibit an in plane 2 × 2 superlattice~\cite{PhysRevB.104.195132}. Initial observation of 2 × 2 superlattice by STM was explained in terms of SD distortion~\cite{luo2021distinct,PhysRevX.11.031026}. On the other hand,  DFT calculations show that ISD distortion is more energetically favored and can be consistent with STM results~\cite{PhysRevB.104.195132}. More recently, Xray diffraction and Raman spectroscopy measurements reported a 2 × 2 × 4 modulation, with one layer of ISD and three consecutive layers of SD, highlighting nontrivial interlayer ordering along the $c$-axis direction~\cite{PhysRevX.11.041030,Wang_2021}. 

By introducing a periodicity in real space that is different from the native lattice, the CDW transition can have significant impact on Fermi surface and topological properties. Quantum oscillation measurements have been used to infer the Fermi surface of AV$_3$Sb$_5$, in particular CsV$_3$Sb$_5$, in the CDW phase~\cite{PhysRevX.11.041030,https://doi.org/10.48550/arxiv.2202.08570,shrestha2022nontrivial,PhysRevLett.127.207002, https://doi.org/10.48550/arxiv.2202.08570,PhysRevB.104.L041103,https://doi.org/10.48550/arxiv.2110.13085}. Consistent with the complexity of the Fermi surface, multiple frequencies were observed in CsV$_3$Sb$_5$ that have been explained based on either SD or ISD distortion; however, no evidence has been observed to show the effects of alternating SD and ISD layers on the Fermi surface~\cite{PhysRevX.11.041030,PhysRevLett.127.207002}. In this experiment, we investigated the Fermi surface of CsV$_3$Sb$_5$ using tunnel diode oscillator (TDO) frequency measurements under applied magnetic fields up to 41.5~T. Our data exhibit quantum oscillations with the largest number of frequencies reported so far, several of which have never been reported. The number of frequencies can not be explained by DFT calculations when only SD or ISD distortion is considered~\cite{PhysRevX.11.041030,PhysRevLett.127.207002}. Instead, our results are consistent with calculations based on 2 × 2 × 4 superlattice, providing stong evidence of the impact of interlayer ordering on the Fermi surface properties.

\begin{figure*}[ht]
    \includegraphics[width=1\linewidth]{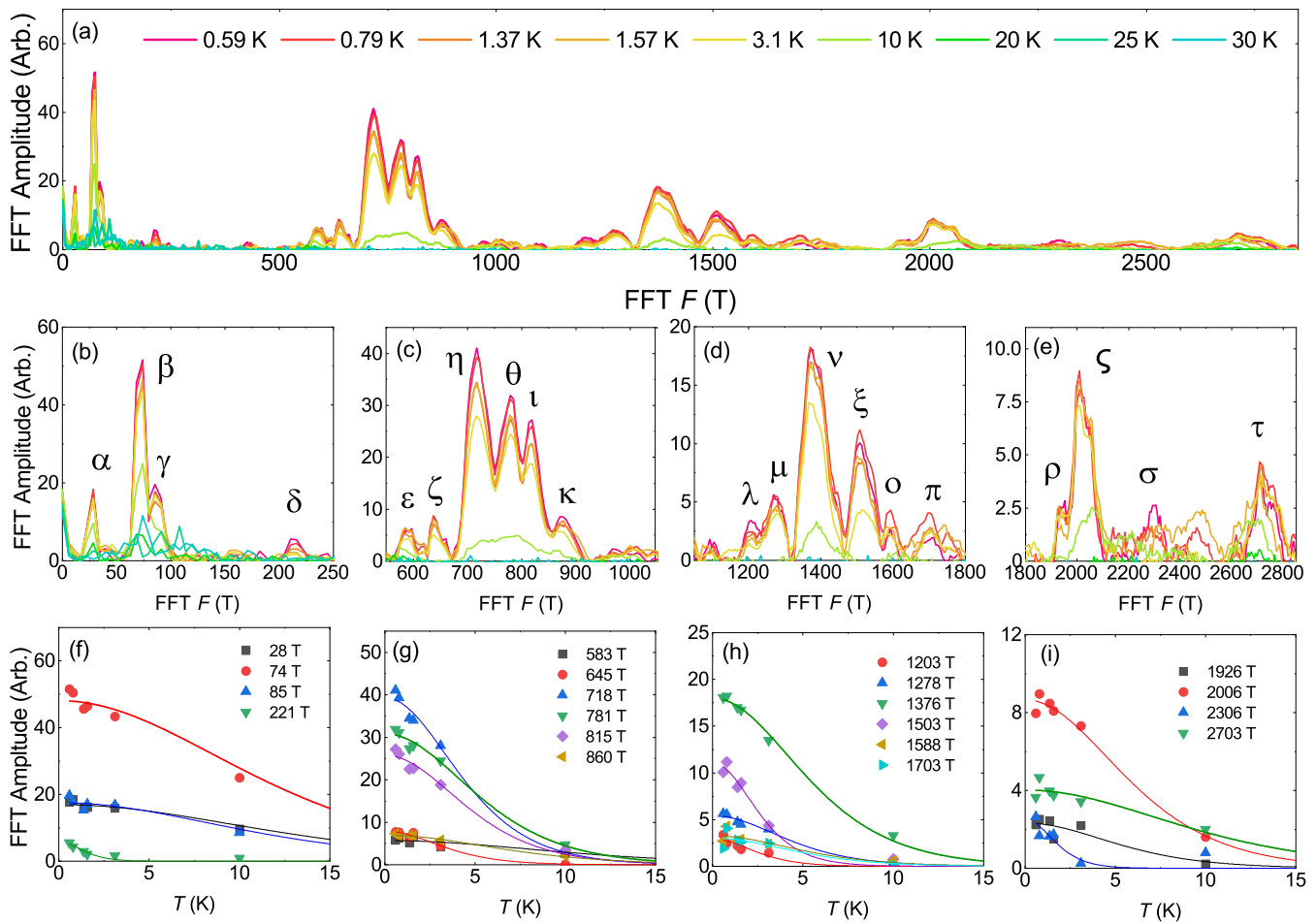}
    \caption{\textbf{(a)} The FFT spectrum plotted up to 2850~T, capturing data from 0.59 K to 30 K in field up to 41.5~T. A noise sampling from 250-450~T was averaged and subtracted from the FFT Spectrum to aid in accuracy of effective mass fittings.  \textbf{(b-e)} The FFT Spectrum is split up into four sub-figures (\textbf{(b)} 0-250~T, \textbf{(c)} 500-1050~T, \textbf{(d)} 1050-1800~T, \textbf{(e)} 1800-2850~T) for better resolution of the peaks. The log of all peaks can be found in Supplementary Table I, where they are compared to recent quantum oscillation articles. \textbf{(f-i)} The experimental data and LK fittings, corresponding to the sub-figure above it (ie. \textbf{(b)} and \textbf{(f)}), are plotted as a function of temperature. The effective mass is calculated from the LK formula (equation in the main text) to determine the effective mass of each orbit. Calculations of $k_F$, $S_F$, and $v_F$ can be found in Supplementary Table II.  }
\end{figure*} 

Figure 1a shows the fast Fourier transform (FFT) spectrum of the TDO data. In the frequency range of 0 - 2850~T, we observed more than twenty distinct frequencies, which is the largest number of frequencies observed in a single sample with a single measurement technique. To assist the discussions, we divide all the frequencies into four groups: 0 - 250~T range (Figure 1b), 500 - 1000~T range (Figure 1c), 1000 - 2000~T range (Figure 1d), and above 2000~T (Figure 1e). Some of the frequencies have been observed in previous studies, which are in good agreement with our results~\cite{PhysRevX.11.041030,https://doi.org/10.48550/arxiv.2202.08570,shrestha2022nontrivial,PhysRevLett.127.207002, https://doi.org/10.48550/arxiv.2202.08570,PhysRevB.104.L041103,https://doi.org/10.48550/arxiv.2110.13085}. We observed seven new frequencies, potentially due to the good sample quality and high sensitivity of the measurement technique. Previous studies mainly have used magnetoresistance while we used TDO, which has been widely used to investigate the Fermi surface of quantum materials with high sensitivity. 

In the low frequency range, we observed four distinct frequencies, $F_{\alpha}$ = 28~T, $F_{\beta}$ = 74~T, $F_{\gamma}$ = 85~T, and $F_{\delta}$ = 221~T. The low frequencies correspond to the small Fermi surfaces, since the the frequency of quantum oscillations is proportional to the area of the extremal orbits. Some of these Fermi surfaces have nontrivial topological properties based on theoretical calculations, confirmed by previous quantum oscillation studies~\cite{shrestha2022nontrivial,PhysRevLett.127.207002}. While $F_{\delta}$ has been previously reported with non-zero $\phi$~\cite{shrestha2022nontrivial}, the signal is quickly damped below the noise resolution.  To investigate the topological nature, Landau level fan diagrams are constructed for $F_{\alpha}$, $F_{\beta}$, and $F_{\gamma}$(Figure 2b). After using filters to isolate a given frequency, Landau levels are labeled corresponding to half integer levels for maximums, and integer filling at minimums~\cite{PhysRevLett.104.086403}. Frequencies obtained from the slope of the Landau level fan diagrams are in good agreement with the frequencies obtained from the Fourier transform, indicating the signals are well preserved when filters are applied. Non-zero phase is revealed by linear extrapolation to zero in 1/$H$ for all three frequencies, close to the predicted $\phi = 0.5$ for topological orbits~\cite{PhysRevLett.82.2147,Fuchs_2010}. It is notable that we observe the 1st Landau level for the 28~T orbit at fields of 41.5~T, making the liner extrapolation highly reliable. Nontrivial $\phi$ has been reported for $F_{\alpha}$ and $F_{\beta}$, with values consistent with ours~\cite{shrestha2022nontrivial,PhysRevLett.127.207002}. Our observation of nontrivial $\phi$ for $F_{\gamma}$ further confirm the richness of nontrivial band topology of CsV$_3$Sb$_5$. 

\begin{figure}[t]
    \includegraphics[width=1\linewidth]{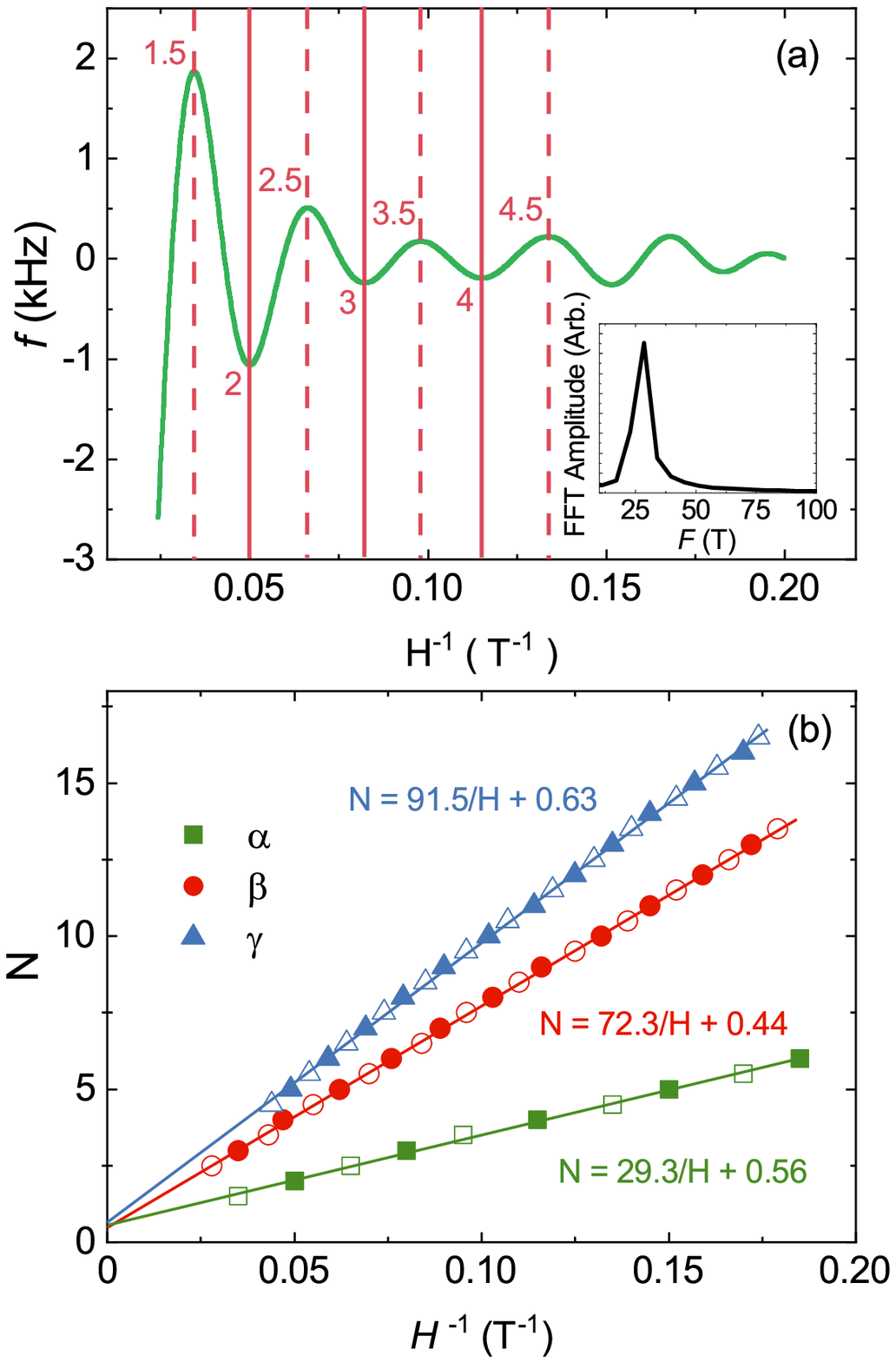}
        \caption{\textbf{(a)} By isolating the 28 T frequency with a high pass filter of 25 and a low pass of 35, the $F_{\alpha}$ orbit Landau levels are able to be assigned (labeled with red text). The inset displays the FFT spectrum for the plotted data, confirming the frequency is well isolated. \textbf{(b)} The Landau level fan diagram for orbits $F_\alpha$, $F_\beta$, and $F_\gamma$ are constructed from the filtered data. By assigning maximums to $n+1/2$ and minimums to $n$, the phase shift, $\phi$, is identified with linear extrapolation to 0 T$^{-1}$~\cite{PhysRevLett.104.086403}. }
\end{figure}

In the range of 500 - 1000 T, we observed six well resolved frequencies, while only up to two frequencies were observed together by a single measurement in previous studies~\cite{PhysRevX.11.041030,shrestha2022nontrivial,PhysRevLett.127.207002, https://doi.org/10.48550/arxiv.2202.08570}. Out of the six orbits, $F_\zeta$ = 646~T and $F_\iota$ = 815~T have never been reported before. Figure 4a shows the TDO frequency data after filtering out frequencies outside the range of 500 - 1000 T. Beating, due to the multiple frequencies of similar magnitude, is clearly seen. The superposition of multiple oscillation waves causes a broadening and shift of the oscillation extrema, making it difficult to assign the Landau level index~\cite{hu2017nearly}. Previous DFT calculations indicate that the frequencies in this range can be understood in terms of structure distortion due to the CDW order~\cite{PhysRevLett.127.207002,PhysRevX.11.041030}. In the native Kagome structure, vanadium $d$ band forms a triangular orbit at the $K$ points at $k_z = 0.5$, giving rise to frequency of 700~T. Upon CDW transition, a structure distortion occurs that changes the Fermi surface~\cite{Zhao_2021,PhysRevB.104.195132}. The Kagome breathing mode can give rise to two types of distortions, SD and its inverse structure ISD~\cite{Zhao_2021,PhysRevB.104.195132}. On the $k_z = 0$ plane, a large V orbit reconstructs into small orbits, while generating a new triangular orbit around the K points~\cite{PhysRevX.11.041030,PhysRevLett.127.207002}. This new triangular orbit gives rise to another frequency, in the range of 700 - 900~T for both distorted structures. From the current understanding, the native triangular orbit from V $d$ band at $K$ points remains upon structure distortion on the $k_z = 0$ plane. Therefore, when either SD or ISD type of distortion is considered, two frequencies between 500 and 1000~T are predicted~\cite{PhysRevX.11.041030,PhysRevLett.127.207002}. All the previous quantum oscillation studies report up to two frequencies in this range, even though not always the same two frequencies are reported by different studies, leading to the conclusion that CDW transition is accompanied by either SD or ISD distortion~\cite{PhysRevX.11.041030,https://doi.org/10.48550/arxiv.2202.08570,shrestha2022nontrivial,PhysRevLett.127.207002, https://doi.org/10.48550/arxiv.2202.08570}.  

Six frequencies clearly observed in this study begs the question to the source of these additional orbits. In the current study, we applied a magnetic field up to 41.5~T, while a maximum of 35~T was applied in the previous studies~\cite{shrestha2022nontrivial}. However, even if we use data only up to 35~T, we get the same six frequencies, as shown in Figure ~3b, indicating the additional frequencies we observed are not due to the magnetic breakdown in higher magnetic field. We propose that our results could be reconciled, when more complicated structure distortion is considered. In reality, both SD and ISD type of structure distortion could exist simultaneously, with alternative SD and ISD distortion from layer to layer. This could result in a distortion with 2 × 2 × 2 modulation, which has been investigated in AV$_3$Sb$_5$~\cite{PhysRevX.11.031050,PhysRevMaterials.5.L111801, PhysRevX.11.031026, Christensen_2021}. Xray diffraction measurement reported more complicated distortions, with 2 × 2 × 4 modulation, i.e., one layer of ISD and three consecutive layers of SD~\cite{PhysRevX.11.041030}. Recent Raman spectroscopy measurement confirmed the 2 × 2 × 4 modulation~\cite{PhysRevX.11.031050}. 

With additional interlayer ordering along the $c$-axis, Fermi surface will naturally be more complicated. At the minimum, two new triangular orbits potentially appear at the K points on the $k_z = 0$ plane, each from one type of distortion~\cite{PhysRevX.11.041030,PhysRevLett.127.207002}. In addition, as the unit cell is enlarged, the size of Brillouin zone will be reduced. Band folding will substantially modify the Fermi surface, leading to more complicated structures~\cite{https://doi.org/10.48550/arxiv.2202.08570,PhysRevB.105.045135}. Such complicated Fermi surface and the resulting quantum oscillation frequencies have been revealed by recent DFT calculations based on 2 × 2 × 4 structure with interlayer ordering of SD and ISD layers~\cite{https://doi.org/10.48550/arxiv.2202.08570}. Ten frequencies are predicted in the range of 400 - 1000~T, in contrast to two frequencies when only SD or ISD distortion is considered~\cite{PhysRevX.11.041030,https://doi.org/10.48550/arxiv.2202.08570,PhysRevLett.127.207002}. Based on the effective mass obtained from the LK fitting to our data, three frequencies with relatively high effective masses would not match our observations. The six frequencies we observed could correspond to six of the remaining seven frequencies predicted (Supplementary Table III). While the predicted frequencies are about 100~T lower than the observed values, it has been pointed out that the calculated quantum oscillation frequencies are very sensitive to the Fermi energy~\cite{PhysRevX.11.041030}. The true value of the Fermi energy in the materials could be different from the value in the calculations, leading to discrepancy between calculated frequencies and the observed ones. In the previous studies, the discrepancy is even larger when only SD or ISD distorted in considered~\cite{PhysRevX.11.041030,PhysRevLett.127.207002}. The qualitative match of our results to the prediction based on 2 × 2 × 4 structure in this frequency range strongly indicate that the Fermi surface of CsV$_3$Sb$_5$ is substantially modified by the interlayer stacking of SD and ISD layers~\cite{https://doi.org/10.48550/arxiv.2202.08570}. Note that we can not rule out distortions with other types of interlayer ordering, e.g., 2 × 2 × 2 structure, as they also give rise to enlarged unit cell leading to more complicated Fermi surface~\cite{Christensen_2021}. Detailed calculations of quantum oscillation frequencies based on the 2 × 2 × 2 structure is needed to investigate the exact periodicity along the $c$ axis in the CDW phase. 

\begin{figure}[t]
    \includegraphics[width=1\linewidth]{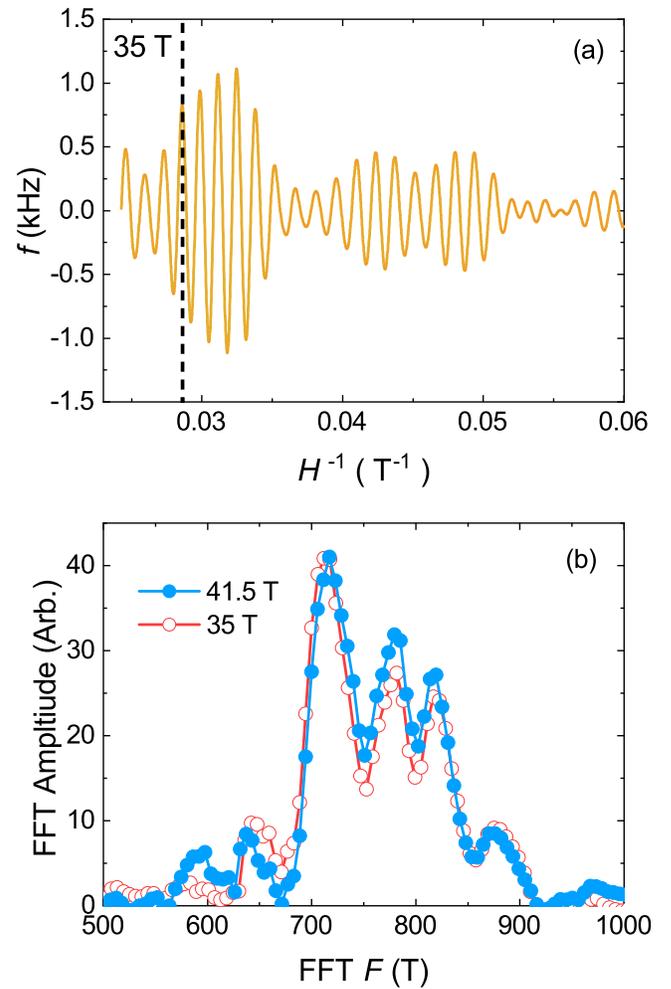}
    \caption{\textbf{(a)} Filters have were used to isolate the mid-range peaks from 500T to 1000 T, and the TDO Frequency for 0.59 K is plotted versus $H^{-1}$. The dashed line indicates the 35~T maximum field from previous studies. \textbf{(b)} The fast Fourier transform of the 0.59 K data plotted from 500 - 1000 T, displayed for the full field range (blue) and for fields limited to 35 T (red).}
\end{figure} 

Another six frequencies are observed in the range of 1200 - 1800~T. At first glance, these six frequencies look like the second harmonics of the six frequencies observed in the range 500 - 1000~T. Careful analysis indicate that the frequencies are not exactly doubled (Supplementary Figure 3). As such, these six frequencies are not the harmonics but independent frequencies corresponding to real orbits on the Fermi surface. Indeed, five out of these six frequencies were reported in previous studies, even though only up two frequencies from the same study~\cite{PhysRevX.11.041030,https://doi.org/10.48550/arxiv.2202.08570,shrestha2022nontrivial,PhysRevLett.127.207002, https://doi.org/10.48550/arxiv.2202.08570}. DFT calculations based on SD or ISD distortion have not revealed any Fermi surface that could potentially corresponds to these frequencies, while eight frequencies are predicted based on 2 × 2 × 4 structure~\cite{https://doi.org/10.48550/arxiv.2202.08570}. Our observation of six frequencies in this range again underscores that the impact of interlayer ordering is pertinent to understanding the Fermi surface properties of CsV$_3$Sb$_5$.

We also observed a few frequencies above 2000~T. However, these frequencies are not predicted even in the 2 × 2 × 4 structure. Similar frequencies were recently reported on thin flakes, but missing in the bulk samples in the same study~\cite{https://doi.org/10.48550/arxiv.2202.08570}. Observation in the thin flakes has been attributed to selective hole doping from oxidation, which potentially reduce Fermi level for certain orbits, increasing the Quantum Oscillation frequency~\cite{https://doi.org/10.48550/arxiv.2202.08570}. Such effects will be more significant in the thin flakes. Since we observed similar frequencies in the bulk sample, the origin of these frequencies may not be related to the oxidation. Further investigation is required to fully understand the Fermi surface properties of CsV$_3$Sb$_5$.

To summarize, our quantum oscillation data provide crucial new insight into the Fermi surface properties of the Kagome superconductor CsV$_3$Sb$_5$. Recent Xray diffraction and Raman spectroscopy measurements have revealed the interlayer ordering in CsV$_3$Sb$_5$ upon CDW transition~\cite{PhysRevX.11.041030,Wang_2021}. The impact of such ordering on the Fermi surface properties have not been previously observed. There have been several quantum oscillation studies of CsV$_3$Sb$_5$, with different frequencies reported~\cite{PhysRevX.11.041030,https://doi.org/10.48550/arxiv.2202.08570,shrestha2022nontrivial,PhysRevLett.127.207002, https://doi.org/10.48550/arxiv.2202.08570,PhysRevB.104.L041103,https://doi.org/10.48550/arxiv.2110.13085}. In the frequency range of 500 - 1000~T, at least four distinct frequencies were previously reported, which already indicate the inconsistency with theoretical calculations when only SD or ISD distortion is considered. However, in the previous studies, only up to two frequencies were observed in a single measurement, leaving the conclusion ambiguous. In the current study, we observed several distinct frequencies in a single measurement that are not predicted when only SD or ISD distortion is considered. Our results provide strong evidence of the interlayer stacking of SD and ISD layers of CsV$_3$Sb$_5$ and the impact on its Fermi surface properties.  

We are highly indebted to Swee Kuan Goh and Wing Chi Yu for making available their band structure calculation results. We would also like to thank Binghai Yan and Li Yang for fruitful discussions. Research at Washington University was supported by McDonnell International Scholars Academy. Research at the National High Magnetic Field Laboratory NHMFL was supported by NSF Cooperative Agreement No.~DMR-1157490, the State of Florida and the DOE. Research at Chinese Academy of Science is supported by grants from the National Natural Science Foundation of China (61888102, 51771224), the National Key Research and Development Projects of China (2018YFA0305800), and the Chinese Academy of Sciences (XDB33030100). 


\bibliography{bib}

\begin{thebibliography}{43}%
\makeatletter
\providecommand \@ifxundefined [1]{%
 \@ifx{#1\undefined}
}%
\providecommand \@ifnum [1]{%
 \ifnum #1\expandafter \@firstoftwo
 \else \expandafter \@secondoftwo
 \fi
}%
\providecommand \@ifx [1]{%
 \ifx #1\expandafter \@firstoftwo
 \else \expandafter \@secondoftwo
 \fi
}%
\providecommand \natexlab [1]{#1}%
\providecommand \enquote  [1]{``#1''}%
\providecommand \bibnamefont  [1]{#1}%
\providecommand \bibfnamefont [1]{#1}%
\providecommand \citenamefont [1]{#1}%
\providecommand \href@noop [0]{\@secondoftwo}%
\providecommand \href [0]{\begingroup \@sanitize@url \@href}%
\providecommand \@href[1]{\@@startlink{#1}\@@href}%
\providecommand \@@href[1]{\endgroup#1\@@endlink}%
\providecommand \@sanitize@url [0]{\catcode `\\12\catcode `\$12\catcode
  `\&12\catcode `\#12\catcode `\^12\catcode `\_12\catcode `\%12\relax}%
\providecommand \@@startlink[1]{}%
\providecommand \@@endlink[0]{}%
\providecommand \url  [0]{\begingroup\@sanitize@url \@url }%
\providecommand \@url [1]{\endgroup\@href {#1}{\urlprefix }}%
\providecommand \urlprefix  [0]{URL }%
\providecommand \Eprint [0]{\href }%
\providecommand \doibase [0]{https://doi.org/}%
\providecommand \selectlanguage [0]{\@gobble}%
\providecommand \bibinfo  [0]{\@secondoftwo}%
\providecommand \bibfield  [0]{\@secondoftwo}%
\providecommand \translation [1]{[#1]}%
\providecommand \BibitemOpen [0]{}%
\providecommand \bibitemStop [0]{}%
\providecommand \bibitemNoStop [0]{.\EOS\space}%
\providecommand \EOS [0]{\spacefactor3000\relax}%
\providecommand \BibitemShut  [1]{\csname bibitem#1\endcsname}%
\let\auto@bib@innerbib\@empty
\bibitem [{\citenamefont {Lin}\ and\ \citenamefont {Nori}(1994)}]{Lin_1994}%
  \BibitemOpen
  \bibfield  {author} {\bibinfo {author} {\bibfnamefont {Y.-L.}\ \bibnamefont
  {Lin}}\ and\ \bibinfo {author} {\bibfnamefont {F.}~\bibnamefont {Nori}},\
  }\bibfield  {title} {\bibinfo {title} {{Quantum interference on the
  Kagome{\textasciiacute} lattice}},\ }\href
  {https://doi.org/10.1103/physrevb.50.15953} {\bibfield  {journal} {\bibinfo
  {journal} {Physical Review B}\ }\textbf {\bibinfo {volume} {50}},\ \bibinfo
  {pages} {15953} (\bibinfo {year} {1994})}\BibitemShut {NoStop}%
\bibitem [{\citenamefont {Li}\ \emph {et~al.}(2018)\citenamefont {Li},
  \citenamefont {Zhuang}, \citenamefont {Wang}, \citenamefont {Feng},
  \citenamefont {Gao}, \citenamefont {Xu}, \citenamefont {Hao}, \citenamefont
  {Wang}, \citenamefont {Zhang}, \citenamefont {Wu}, \citenamefont {Dou},
  \citenamefont {Chen}, \citenamefont {Hu},\ and\ \citenamefont
  {Du}}]{https://doi.org/10.48550/arxiv.1708.04448}%
  \BibitemOpen
  \bibfield  {author} {\bibinfo {author} {\bibfnamefont {Z.}~\bibnamefont
  {Li}}, \bibinfo {author} {\bibfnamefont {J.}~\bibnamefont {Zhuang}}, \bibinfo
  {author} {\bibfnamefont {L.}~\bibnamefont {Wang}}, \bibinfo {author}
  {\bibfnamefont {H.}~\bibnamefont {Feng}}, \bibinfo {author} {\bibfnamefont
  {Q.}~\bibnamefont {Gao}}, \bibinfo {author} {\bibfnamefont {X.}~\bibnamefont
  {Xu}}, \bibinfo {author} {\bibfnamefont {W.}~\bibnamefont {Hao}}, \bibinfo
  {author} {\bibfnamefont {X.}~\bibnamefont {Wang}}, \bibinfo {author}
  {\bibfnamefont {C.}~\bibnamefont {Zhang}}, \bibinfo {author} {\bibfnamefont
  {K.}~\bibnamefont {Wu}}, \bibinfo {author} {\bibfnamefont {S.~X.}\
  \bibnamefont {Dou}}, \bibinfo {author} {\bibfnamefont {L.}~\bibnamefont
  {Chen}}, \bibinfo {author} {\bibfnamefont {Z.}~\bibnamefont {Hu}},\ and\
  \bibinfo {author} {\bibfnamefont {Y.}~\bibnamefont {Du}},\ }\bibfield
  {title} {\bibinfo {title} {{Realization of flat band with possible nontrivial
  topology in electronic Kagome lattice}},\ }\href
  {https://doi.org/10.1126/sciadv.aau4511} {\bibfield  {journal} {\bibinfo
  {journal} {Science Advances}\ }\textbf {\bibinfo {volume} {4}},\ \bibinfo
  {pages} {1343} (\bibinfo {year} {2018})}\BibitemShut {NoStop}%
\bibitem [{\citenamefont {Guo}\ and\ \citenamefont
  {Franz}(2009)}]{PhysRevB.80.113102}%
  \BibitemOpen
  \bibfield  {author} {\bibinfo {author} {\bibfnamefont {H.-M.}\ \bibnamefont
  {Guo}}\ and\ \bibinfo {author} {\bibfnamefont {M.}~\bibnamefont {Franz}},\
  }\bibfield  {title} {\bibinfo {title} {{Topological insulator on the Kagome
  lattice}},\ }\href {https://doi.org/10.1103/PhysRevB.80.113102} {\bibfield
  {journal} {\bibinfo  {journal} {Phys. Rev. B}\ }\textbf {\bibinfo {volume}
  {80}},\ \bibinfo {pages} {113102} (\bibinfo {year} {2009})}\BibitemShut
  {NoStop}%
\bibitem [{\citenamefont {Ortiz}\ \emph {et~al.}(2019)\citenamefont {Ortiz},
  \citenamefont {Gomes}, \citenamefont {Morey}, \citenamefont {Winiarski},
  \citenamefont {Bordelon}, \citenamefont {Mangum}, \citenamefont {Oswald},
  \citenamefont {Rodriguez-Rivera}, \citenamefont {Neilson}, \citenamefont
  {Wilson}, \citenamefont {Ertekin}, \citenamefont {McQueen},\ and\
  \citenamefont {Toberer}}]{PhysRevMaterials.3.094407}%
  \BibitemOpen
  \bibfield  {author} {\bibinfo {author} {\bibfnamefont {B.~R.}\ \bibnamefont
  {Ortiz}}, \bibinfo {author} {\bibfnamefont {L.~C.}\ \bibnamefont {Gomes}},
  \bibinfo {author} {\bibfnamefont {J.~R.}\ \bibnamefont {Morey}}, \bibinfo
  {author} {\bibfnamefont {M.}~\bibnamefont {Winiarski}}, \bibinfo {author}
  {\bibfnamefont {M.}~\bibnamefont {Bordelon}}, \bibinfo {author}
  {\bibfnamefont {J.~S.}\ \bibnamefont {Mangum}}, \bibinfo {author}
  {\bibfnamefont {I.~W.~H.}\ \bibnamefont {Oswald}}, \bibinfo {author}
  {\bibfnamefont {J.~A.}\ \bibnamefont {Rodriguez-Rivera}}, \bibinfo {author}
  {\bibfnamefont {J.~R.}\ \bibnamefont {Neilson}}, \bibinfo {author}
  {\bibfnamefont {S.~D.}\ \bibnamefont {Wilson}}, \bibinfo {author}
  {\bibfnamefont {E.}~\bibnamefont {Ertekin}}, \bibinfo {author} {\bibfnamefont
  {T.~M.}\ \bibnamefont {McQueen}},\ and\ \bibinfo {author} {\bibfnamefont
  {E.~S.}\ \bibnamefont {Toberer}},\ }\bibfield  {title} {\bibinfo {title} {New
  kagome prototype materials: discovery of
  {KV}$_3${Sb}$_5$,{Rb}{V}$_3${Sb}$_5$, and {CsV}$_3${Sb}$_5$},\ }\href
  {https://doi.org/10.1103/PhysRevMaterials.3.094407} {\bibfield  {journal}
  {\bibinfo  {journal} {Phys. Rev. Materials}\ }\textbf {\bibinfo {volume}
  {3}},\ \bibinfo {pages} {094407} (\bibinfo {year} {2019})}\BibitemShut
  {NoStop}%
\bibitem [{\citenamefont {Tan}\ \emph {et~al.}(2021{\natexlab{a}})\citenamefont
  {Tan}, \citenamefont {Liu}, \citenamefont {Wang},\ and\ \citenamefont
  {Yan}}]{Tan_2021}%
  \BibitemOpen
  \bibfield  {author} {\bibinfo {author} {\bibfnamefont {H.}~\bibnamefont
  {Tan}}, \bibinfo {author} {\bibfnamefont {Y.}~\bibnamefont {Liu}}, \bibinfo
  {author} {\bibfnamefont {Z.}~\bibnamefont {Wang}},\ and\ \bibinfo {author}
  {\bibfnamefont {B.}~\bibnamefont {Yan}},\ }\bibfield  {title} {\bibinfo
  {title} {Charge density waves and electronic properties of superconducting
  kagome metals},\ }\href {https://doi.org/10.1103/physrevlett.127.046401}
  {\bibfield  {journal} {\bibinfo  {journal} {Physical Review Letters}\
  }\textbf {\bibinfo {volume} {127}},\ \bibinfo {pages} {046401} (\bibinfo
  {year} {2021}{\natexlab{a}})}\BibitemShut {NoStop}%
\bibitem [{\citenamefont {Ortiz}\ \emph {et~al.}(2020)\citenamefont {Ortiz},
  \citenamefont {Teicher}, \citenamefont {Hu}, \citenamefont {Zuo},
  \citenamefont {Sarte}, \citenamefont {Schueller}, \citenamefont {Abeykoon},
  \citenamefont {Krogstad}, \citenamefont {Rosenkranz}, \citenamefont {Osborn},
  \citenamefont {Seshadri}, \citenamefont {Balents}, \citenamefont {He},\ and\
  \citenamefont {Wilson}}]{PhysRevLett.125.247002}%
  \BibitemOpen
  \bibfield  {author} {\bibinfo {author} {\bibfnamefont {B.~R.}\ \bibnamefont
  {Ortiz}}, \bibinfo {author} {\bibfnamefont {S.~M.~L.}\ \bibnamefont
  {Teicher}}, \bibinfo {author} {\bibfnamefont {Y.}~\bibnamefont {Hu}},
  \bibinfo {author} {\bibfnamefont {J.~L.}\ \bibnamefont {Zuo}}, \bibinfo
  {author} {\bibfnamefont {P.~M.}\ \bibnamefont {Sarte}}, \bibinfo {author}
  {\bibfnamefont {E.~C.}\ \bibnamefont {Schueller}}, \bibinfo {author}
  {\bibfnamefont {A.~M.~M.}\ \bibnamefont {Abeykoon}}, \bibinfo {author}
  {\bibfnamefont {M.~J.}\ \bibnamefont {Krogstad}}, \bibinfo {author}
  {\bibfnamefont {S.}~\bibnamefont {Rosenkranz}}, \bibinfo {author}
  {\bibfnamefont {R.}~\bibnamefont {Osborn}}, \bibinfo {author} {\bibfnamefont
  {R.}~\bibnamefont {Seshadri}}, \bibinfo {author} {\bibfnamefont
  {L.}~\bibnamefont {Balents}}, \bibinfo {author} {\bibfnamefont
  {J.}~\bibnamefont {He}},\ and\ \bibinfo {author} {\bibfnamefont {S.~D.}\
  \bibnamefont {Wilson}},\ }\bibfield  {title} {\bibinfo {title}
  {{CsV}$_{3}${Sb}$_{5}$: A ${Z}_{2}$ {Topological Kagome Metal with a
  Superconducting Ground State}},\ }\href
  {https://doi.org/10.1103/PhysRevLett.125.247002} {\bibfield  {journal}
  {\bibinfo  {journal} {Phys. Rev. Lett.}\ }\textbf {\bibinfo {volume} {125}},\
  \bibinfo {pages} {247002} (\bibinfo {year} {2020})}\BibitemShut {NoStop}%
\bibitem [{\citenamefont {Li}\ \emph {et~al.}(2021)\citenamefont {Li},
  \citenamefont {Zhang}, \citenamefont {Yilmaz}, \citenamefont {Pai},
  \citenamefont {Marvinney}, \citenamefont {Said}, \citenamefont {Yin},
  \citenamefont {Gong}, \citenamefont {Tu}, \citenamefont {Vescovo},
  \citenamefont {Nelson}, \citenamefont {Moore}, \citenamefont {Murakami},
  \citenamefont {Lei}, \citenamefont {Lee}, \citenamefont {Lawrie},\ and\
  \citenamefont {Miao}}]{PhysRevX.11.031050}%
  \BibitemOpen
  \bibfield  {author} {\bibinfo {author} {\bibfnamefont {H.}~\bibnamefont
  {Li}}, \bibinfo {author} {\bibfnamefont {T.~T.}\ \bibnamefont {Zhang}},
  \bibinfo {author} {\bibfnamefont {T.}~\bibnamefont {Yilmaz}}, \bibinfo
  {author} {\bibfnamefont {Y.~Y.}\ \bibnamefont {Pai}}, \bibinfo {author}
  {\bibfnamefont {C.~E.}\ \bibnamefont {Marvinney}}, \bibinfo {author}
  {\bibfnamefont {A.}~\bibnamefont {Said}}, \bibinfo {author} {\bibfnamefont
  {Q.~W.}\ \bibnamefont {Yin}}, \bibinfo {author} {\bibfnamefont {C.~S.}\
  \bibnamefont {Gong}}, \bibinfo {author} {\bibfnamefont {Z.~J.}\ \bibnamefont
  {Tu}}, \bibinfo {author} {\bibfnamefont {E.}~\bibnamefont {Vescovo}},
  \bibinfo {author} {\bibfnamefont {C.~S.}\ \bibnamefont {Nelson}}, \bibinfo
  {author} {\bibfnamefont {R.~G.}\ \bibnamefont {Moore}}, \bibinfo {author}
  {\bibfnamefont {S.}~\bibnamefont {Murakami}}, \bibinfo {author}
  {\bibfnamefont {H.~C.}\ \bibnamefont {Lei}}, \bibinfo {author} {\bibfnamefont
  {H.~N.}\ \bibnamefont {Lee}}, \bibinfo {author} {\bibfnamefont {B.~J.}\
  \bibnamefont {Lawrie}},\ and\ \bibinfo {author} {\bibfnamefont
  {H.}~\bibnamefont {Miao}},\ }\bibfield  {title} {\bibinfo {title}
  {{Observation of Unconventional Charge Density Wave without Acoustic Phonon
  Anomaly in Kagome Superconductors ${A\mathrm{V}}_{3}{\mathrm{Sb}}_{5}$
  ($A=\mathrm{Rb}$, Cs)}},\ }\href {https://doi.org/10.1103/PhysRevX.11.031050}
  {\bibfield  {journal} {\bibinfo  {journal} {Phys. Rev. X}\ }\textbf {\bibinfo
  {volume} {11}},\ \bibinfo {pages} {031050} (\bibinfo {year}
  {2021})}\BibitemShut {NoStop}%
\bibitem [{\citenamefont {Nakayama}\ \emph {et~al.}(2021)\citenamefont
  {Nakayama}, \citenamefont {Li}, \citenamefont {Kato}, \citenamefont {Liu},
  \citenamefont {Wang}, \citenamefont {Takahashi}, \citenamefont {Yao},\ and\
  \citenamefont {Sato}}]{PhysRevB.104.L161112}%
  \BibitemOpen
  \bibfield  {author} {\bibinfo {author} {\bibfnamefont {K.}~\bibnamefont
  {Nakayama}}, \bibinfo {author} {\bibfnamefont {Y.}~\bibnamefont {Li}},
  \bibinfo {author} {\bibfnamefont {T.}~\bibnamefont {Kato}}, \bibinfo {author}
  {\bibfnamefont {M.}~\bibnamefont {Liu}}, \bibinfo {author} {\bibfnamefont
  {Z.}~\bibnamefont {Wang}}, \bibinfo {author} {\bibfnamefont {T.}~\bibnamefont
  {Takahashi}}, \bibinfo {author} {\bibfnamefont {Y.}~\bibnamefont {Yao}},\
  and\ \bibinfo {author} {\bibfnamefont {T.}~\bibnamefont {Sato}},\ }\bibfield
  {title} {\bibinfo {title} {{Multiple energy scales and anisotropic energy gap
  in the charge-density-wave phase of the Kagome superconductor
  ${\mathrm{CsV}}_{3}{\mathrm{Sb}}_{5}$}},\ }\href
  {https://doi.org/10.1103/PhysRevB.104.L161112} {\bibfield  {journal}
  {\bibinfo  {journal} {Phys. Rev. B}\ }\textbf {\bibinfo {volume} {104}},\
  \bibinfo {pages} {L161112} (\bibinfo {year} {2021})}\BibitemShut {NoStop}%
\bibitem [{\citenamefont {Cho}\ \emph {et~al.}(2021)\citenamefont {Cho},
  \citenamefont {Ma}, \citenamefont {Xia}, \citenamefont {Yang}, \citenamefont
  {Liu}, \citenamefont {Huang}, \citenamefont {Jiang}, \citenamefont {Lu},
  \citenamefont {Liu}, \citenamefont {Liu}, \citenamefont {Li}, \citenamefont
  {Wang}, \citenamefont {Liu}, \citenamefont {Jia}, \citenamefont {Guo},
  \citenamefont {Liu},\ and\ \citenamefont {Shen}}]{PhysRevLett.127.236401}%
  \BibitemOpen
  \bibfield  {author} {\bibinfo {author} {\bibfnamefont {S.}~\bibnamefont
  {Cho}}, \bibinfo {author} {\bibfnamefont {H.}~\bibnamefont {Ma}}, \bibinfo
  {author} {\bibfnamefont {W.}~\bibnamefont {Xia}}, \bibinfo {author}
  {\bibfnamefont {Y.}~\bibnamefont {Yang}}, \bibinfo {author} {\bibfnamefont
  {Z.}~\bibnamefont {Liu}}, \bibinfo {author} {\bibfnamefont {Z.}~\bibnamefont
  {Huang}}, \bibinfo {author} {\bibfnamefont {Z.}~\bibnamefont {Jiang}},
  \bibinfo {author} {\bibfnamefont {X.}~\bibnamefont {Lu}}, \bibinfo {author}
  {\bibfnamefont {J.}~\bibnamefont {Liu}}, \bibinfo {author} {\bibfnamefont
  {Z.}~\bibnamefont {Liu}}, \bibinfo {author} {\bibfnamefont {J.}~\bibnamefont
  {Li}}, \bibinfo {author} {\bibfnamefont {J.}~\bibnamefont {Wang}}, \bibinfo
  {author} {\bibfnamefont {Y.}~\bibnamefont {Liu}}, \bibinfo {author}
  {\bibfnamefont {J.}~\bibnamefont {Jia}}, \bibinfo {author} {\bibfnamefont
  {Y.}~\bibnamefont {Guo}}, \bibinfo {author} {\bibfnamefont {J.}~\bibnamefont
  {Liu}},\ and\ \bibinfo {author} {\bibfnamefont {D.}~\bibnamefont {Shen}},\
  }\bibfield  {title} {\bibinfo {title} {{Emergence of New van Hove
  Singularities in the Charge Density Wave State of a Topological Kagome Metal
  ${\mathrm{RbV}}_{3}{\mathrm{Sb}}_{5}$}},\ }\href
  {https://doi.org/10.1103/PhysRevLett.127.236401} {\bibfield  {journal}
  {\bibinfo  {journal} {Phys. Rev. Lett.}\ }\textbf {\bibinfo {volume} {127}},\
  \bibinfo {pages} {236401} (\bibinfo {year} {2021})}\BibitemShut {NoStop}%
\bibitem [{\citenamefont {Zhao}\ \emph
  {et~al.}(2021{\natexlab{a}})\citenamefont {Zhao}, \citenamefont {Wu},
  \citenamefont {Wang},\ and\ \citenamefont {Yang}}]{PhysRevB.103.L241117}%
  \BibitemOpen
  \bibfield  {author} {\bibinfo {author} {\bibfnamefont {J.}~\bibnamefont
  {Zhao}}, \bibinfo {author} {\bibfnamefont {W.}~\bibnamefont {Wu}}, \bibinfo
  {author} {\bibfnamefont {Y.}~\bibnamefont {Wang}},\ and\ \bibinfo {author}
  {\bibfnamefont {S.~A.}\ \bibnamefont {Yang}},\ }\bibfield  {title} {\bibinfo
  {title} {{Electronic correlations in the normal state of the kagome
  superconductor ${\mathrm{KV}}_{3}{\mathrm{Sb}}_{5}$}},\ }\href
  {https://doi.org/10.1103/PhysRevB.103.L241117} {\bibfield  {journal}
  {\bibinfo  {journal} {Phys. Rev. B}\ }\textbf {\bibinfo {volume} {103}},\
  \bibinfo {pages} {L241117} (\bibinfo {year}
  {2021}{\natexlab{a}})}\BibitemShut {NoStop}%
\bibitem [{\citenamefont {Liang}\ \emph
  {et~al.}(2021{\natexlab{a}})\citenamefont {Liang}, \citenamefont {Hou},
  \citenamefont {Zhang}, \citenamefont {Ma}, \citenamefont {Wu}, \citenamefont
  {Zhang}, \citenamefont {Yu}, \citenamefont {Ying}, \citenamefont {Jiang},
  \citenamefont {Shan}, \citenamefont {Wang},\ and\ \citenamefont
  {Chen}}]{Liang_2021}%
  \BibitemOpen
  \bibfield  {author} {\bibinfo {author} {\bibfnamefont {Z.}~\bibnamefont
  {Liang}}, \bibinfo {author} {\bibfnamefont {X.}~\bibnamefont {Hou}}, \bibinfo
  {author} {\bibfnamefont {F.}~\bibnamefont {Zhang}}, \bibinfo {author}
  {\bibfnamefont {W.}~\bibnamefont {Ma}}, \bibinfo {author} {\bibfnamefont
  {P.}~\bibnamefont {Wu}}, \bibinfo {author} {\bibfnamefont {Z.}~\bibnamefont
  {Zhang}}, \bibinfo {author} {\bibfnamefont {F.}~\bibnamefont {Yu}}, \bibinfo
  {author} {\bibfnamefont {J.-J.}\ \bibnamefont {Ying}}, \bibinfo {author}
  {\bibfnamefont {K.}~\bibnamefont {Jiang}}, \bibinfo {author} {\bibfnamefont
  {L.}~\bibnamefont {Shan}}, \bibinfo {author} {\bibfnamefont {Z.}~\bibnamefont
  {Wang}},\ and\ \bibinfo {author} {\bibfnamefont {X.-H.}\ \bibnamefont
  {Chen}},\ }\bibfield  {title} {\bibinfo {title} {{Three-Dimensional Charge
  Density Wave and Surface-Dependent Vortex-Core States in a Kagome
  Superconductor}},\ }\href {https://doi.org/10.1103/physrevx.11.031026}
  {\bibfield  {journal} {\bibinfo  {journal} {Physical Review X}\ }\textbf
  {\bibinfo {volume} {11}},\ \bibinfo {pages} {031026} (\bibinfo {year}
  {2021}{\natexlab{a}})}\BibitemShut {NoStop}%
\bibitem [{\citenamefont {Wang}\ \emph {et~al.}(2020)\citenamefont {Wang},
  \citenamefont {Yang}, \citenamefont {Sivakumar}, \citenamefont {Ortiz},
  \citenamefont {Teicher}, \citenamefont {Wu}, \citenamefont {Srivastava},
  \citenamefont {Garg}, \citenamefont {Liu}, \citenamefont {Parkin},
  \citenamefont {Toberer}, \citenamefont {McQueen}, \citenamefont {Wilson},\
  and\ \citenamefont {Ali}}]{https://doi.org/10.48550/arxiv.2012.05898}%
  \BibitemOpen
  \bibfield  {author} {\bibinfo {author} {\bibfnamefont {Y.}~\bibnamefont
  {Wang}}, \bibinfo {author} {\bibfnamefont {S.}~\bibnamefont {Yang}}, \bibinfo
  {author} {\bibfnamefont {P.~K.}\ \bibnamefont {Sivakumar}}, \bibinfo {author}
  {\bibfnamefont {B.~R.}\ \bibnamefont {Ortiz}}, \bibinfo {author}
  {\bibfnamefont {S.~M.~L.}\ \bibnamefont {Teicher}}, \bibinfo {author}
  {\bibfnamefont {H.}~\bibnamefont {Wu}}, \bibinfo {author} {\bibfnamefont
  {A.~K.}\ \bibnamefont {Srivastava}}, \bibinfo {author} {\bibfnamefont
  {C.}~\bibnamefont {Garg}}, \bibinfo {author} {\bibfnamefont {D.}~\bibnamefont
  {Liu}}, \bibinfo {author} {\bibfnamefont {S.~S.~P.}\ \bibnamefont {Parkin}},
  \bibinfo {author} {\bibfnamefont {E.~S.}\ \bibnamefont {Toberer}}, \bibinfo
  {author} {\bibfnamefont {T.}~\bibnamefont {McQueen}}, \bibinfo {author}
  {\bibfnamefont {S.~D.}\ \bibnamefont {Wilson}},\ and\ \bibinfo {author}
  {\bibfnamefont {M.~N.}\ \bibnamefont {Ali}},\ }\bibfield  {title} {\bibinfo
  {title} {{Proximity-induced spin-triplet superconductivity and edge
  supercurrent in the topological Kagome metal, $\mathrm{K_{1-x}V_3Sb_5}$}},\
  }\href {https://doi.org/10.48550/ARXIV.2012.05898} {\bibfield  {journal}
  {\bibinfo  {journal} {arXiV}\ ,\ \bibinfo {pages} {05898}} (\bibinfo {year}
  {2020})}\BibitemShut {NoStop}%
\bibitem [{\citenamefont {Wang}\ \emph
  {et~al.}(2021{\natexlab{a}})\citenamefont {Wang}, \citenamefont {Ma},
  \citenamefont {Zhang}, \citenamefont {Yang}, \citenamefont {Zhao},
  \citenamefont {Ou}, \citenamefont {Zhu}, \citenamefont {Ni}, \citenamefont
  {Lu}, \citenamefont {Chen}, \citenamefont {Jiang}, \citenamefont {Yu},
  \citenamefont {Zhang}, \citenamefont {Dong}, \citenamefont {Hu},
  \citenamefont {Gao},\ and\ \citenamefont
  {Zhao}}]{https://doi.org/10.48550/arxiv.2104.05556}%
  \BibitemOpen
  \bibfield  {author} {\bibinfo {author} {\bibfnamefont {Z.}~\bibnamefont
  {Wang}}, \bibinfo {author} {\bibfnamefont {S.}~\bibnamefont {Ma}}, \bibinfo
  {author} {\bibfnamefont {Y.}~\bibnamefont {Zhang}}, \bibinfo {author}
  {\bibfnamefont {H.}~\bibnamefont {Yang}}, \bibinfo {author} {\bibfnamefont
  {Z.}~\bibnamefont {Zhao}}, \bibinfo {author} {\bibfnamefont {Y.}~\bibnamefont
  {Ou}}, \bibinfo {author} {\bibfnamefont {Y.}~\bibnamefont {Zhu}}, \bibinfo
  {author} {\bibfnamefont {S.}~\bibnamefont {Ni}}, \bibinfo {author}
  {\bibfnamefont {Z.}~\bibnamefont {Lu}}, \bibinfo {author} {\bibfnamefont
  {H.}~\bibnamefont {Chen}}, \bibinfo {author} {\bibfnamefont {K.}~\bibnamefont
  {Jiang}}, \bibinfo {author} {\bibfnamefont {L.}~\bibnamefont {Yu}}, \bibinfo
  {author} {\bibfnamefont {Y.}~\bibnamefont {Zhang}}, \bibinfo {author}
  {\bibfnamefont {X.}~\bibnamefont {Dong}}, \bibinfo {author} {\bibfnamefont
  {J.}~\bibnamefont {Hu}}, \bibinfo {author} {\bibfnamefont {H.-J.}\
  \bibnamefont {Gao}},\ and\ \bibinfo {author} {\bibfnamefont {Z.}~\bibnamefont
  {Zhao}},\ }\bibfield  {title} {\bibinfo {title} {{Distinctive momentum
  dependent charge-density-wave gap observed in CsV$_3$Sb$_5$ superconductor
  with topological Kagome lattice}},\ }\href
  {https://doi.org/10.48550/ARXIV.2104.05556} {\bibfield  {journal} {\bibinfo
  {journal} {arXiv}\ }\textbf {\bibinfo {volume} {2104}},\ \bibinfo {pages}
  {05556} (\bibinfo {year} {2021}{\natexlab{a}})}\BibitemShut {NoStop}%
\bibitem [{\citenamefont {Qian}\ \emph {et~al.}(2021)\citenamefont {Qian},
  \citenamefont {Christensen}, \citenamefont {Hu}, \citenamefont {Saha},
  \citenamefont {Andersen}, \citenamefont {Fernandes}, \citenamefont {Birol},\
  and\ \citenamefont {Ni}}]{Qian_2021}%
  \BibitemOpen
  \bibfield  {author} {\bibinfo {author} {\bibfnamefont {T.}~\bibnamefont
  {Qian}}, \bibinfo {author} {\bibfnamefont {M.~H.}\ \bibnamefont
  {Christensen}}, \bibinfo {author} {\bibfnamefont {C.}~\bibnamefont {Hu}},
  \bibinfo {author} {\bibfnamefont {A.}~\bibnamefont {Saha}}, \bibinfo {author}
  {\bibfnamefont {B.~M.}\ \bibnamefont {Andersen}}, \bibinfo {author}
  {\bibfnamefont {R.~M.}\ \bibnamefont {Fernandes}}, \bibinfo {author}
  {\bibfnamefont {T.}~\bibnamefont {Birol}},\ and\ \bibinfo {author}
  {\bibfnamefont {N.}~\bibnamefont {Ni}},\ }\bibfield  {title} {\bibinfo
  {title} {{Revealing the competition between charge density wave and
  superconductivity in CsV$_3$Sb$_5$ through uniaxial strain}},\ }\href
  {https://doi.org/10.1103/physrevb.104.144506} {\bibfield  {journal} {\bibinfo
   {journal} {Physical Review B}\ }\textbf {\bibinfo {volume} {104}},\ \bibinfo
  {pages} {144506} (\bibinfo {year} {2021})}\BibitemShut {NoStop}%
\bibitem [{\citenamefont {Yu}\ \emph {et~al.}(2021{\natexlab{a}})\citenamefont
  {Yu}, \citenamefont {Ma}, \citenamefont {Zhuo}, \citenamefont {Liu},
  \citenamefont {Wen}, \citenamefont {Lei}, \citenamefont {Ying},\ and\
  \citenamefont {Chen}}]{Yu_2021}%
  \BibitemOpen
  \bibfield  {author} {\bibinfo {author} {\bibfnamefont {F.~H.}\ \bibnamefont
  {Yu}}, \bibinfo {author} {\bibfnamefont {D.~H.}\ \bibnamefont {Ma}}, \bibinfo
  {author} {\bibfnamefont {W.~Z.}\ \bibnamefont {Zhuo}}, \bibinfo {author}
  {\bibfnamefont {S.~Q.}\ \bibnamefont {Liu}}, \bibinfo {author} {\bibfnamefont
  {X.~K.}\ \bibnamefont {Wen}}, \bibinfo {author} {\bibfnamefont
  {B.}~\bibnamefont {Lei}}, \bibinfo {author} {\bibfnamefont {J.~J.}\
  \bibnamefont {Ying}},\ and\ \bibinfo {author} {\bibfnamefont {X.~H.}\
  \bibnamefont {Chen}},\ }\bibfield  {title} {\bibinfo {title} {Unusual
  competition of superconductivity and charge-density-wave state in a
  compressed topological kagome metal},\ }\href
  {https://doi.org/10.1038/s41467-021-23928-w} {\bibfield  {journal} {\bibinfo
  {journal} {Nature Communications}\ }\textbf {\bibinfo {volume} {12}},\
  \bibinfo {pages} {3645} (\bibinfo {year} {2021}{\natexlab{a}})}\BibitemShut
  {NoStop}%
\bibitem [{\citenamefont {Chen}\ \emph
  {et~al.}(2021{\natexlab{a}})\citenamefont {Chen}, \citenamefont {Wang},
  \citenamefont {Yin}, \citenamefont {Gu}, \citenamefont {Jiang}, \citenamefont
  {Tu}, \citenamefont {Gong}, \citenamefont {Uwatoko}, \citenamefont {Sun},
  \citenamefont {Lei}, \citenamefont {Hu},\ and\ \citenamefont
  {Cheng}}]{PhysRevLett.126.247001}%
  \BibitemOpen
  \bibfield  {author} {\bibinfo {author} {\bibfnamefont {K.~Y.}\ \bibnamefont
  {Chen}}, \bibinfo {author} {\bibfnamefont {N.~N.}\ \bibnamefont {Wang}},
  \bibinfo {author} {\bibfnamefont {Q.~W.}\ \bibnamefont {Yin}}, \bibinfo
  {author} {\bibfnamefont {Y.~H.}\ \bibnamefont {Gu}}, \bibinfo {author}
  {\bibfnamefont {K.}~\bibnamefont {Jiang}}, \bibinfo {author} {\bibfnamefont
  {Z.~J.}\ \bibnamefont {Tu}}, \bibinfo {author} {\bibfnamefont {C.~S.}\
  \bibnamefont {Gong}}, \bibinfo {author} {\bibfnamefont {Y.}~\bibnamefont
  {Uwatoko}}, \bibinfo {author} {\bibfnamefont {J.~P.}\ \bibnamefont {Sun}},
  \bibinfo {author} {\bibfnamefont {H.~C.}\ \bibnamefont {Lei}}, \bibinfo
  {author} {\bibfnamefont {J.~P.}\ \bibnamefont {Hu}},\ and\ \bibinfo {author}
  {\bibfnamefont {J.-G.}\ \bibnamefont {Cheng}},\ }\bibfield  {title} {\bibinfo
  {title} {{Double Superconducting Dome and Triple Enhancement of ${T}_{c}$ in
  the Kagome Superconductor ${\mathrm{CsV}}_{3}{\mathrm{Sb}}_{5}$ under High
  Pressure}},\ }\href {https://doi.org/10.1103/PhysRevLett.126.247001}
  {\bibfield  {journal} {\bibinfo  {journal} {Phys. Rev. Lett.}\ }\textbf
  {\bibinfo {volume} {126}},\ \bibinfo {pages} {247001} (\bibinfo {year}
  {2021}{\natexlab{a}})}\BibitemShut {NoStop}%
\bibitem [{\citenamefont {Wang}\ \emph
  {et~al.}(2021{\natexlab{b}})\citenamefont {Wang}, \citenamefont {Kong},
  \citenamefont {Shi}, \citenamefont {Pei}, \citenamefont {Wen}, \citenamefont
  {Gao}, \citenamefont {Zhao}, \citenamefont {Yin}, \citenamefont {Wu},
  \citenamefont {Li}, \citenamefont {Lei}, \citenamefont {Li}, \citenamefont
  {Chen}, \citenamefont {Yan},\ and\ \citenamefont {Qi}}]{Wang_2021}%
  \BibitemOpen
  \bibfield  {author} {\bibinfo {author} {\bibfnamefont {Q.}~\bibnamefont
  {Wang}}, \bibinfo {author} {\bibfnamefont {P.}~\bibnamefont {Kong}}, \bibinfo
  {author} {\bibfnamefont {W.}~\bibnamefont {Shi}}, \bibinfo {author}
  {\bibfnamefont {C.}~\bibnamefont {Pei}}, \bibinfo {author} {\bibfnamefont
  {C.}~\bibnamefont {Wen}}, \bibinfo {author} {\bibfnamefont {L.}~\bibnamefont
  {Gao}}, \bibinfo {author} {\bibfnamefont {Y.}~\bibnamefont {Zhao}}, \bibinfo
  {author} {\bibfnamefont {Q.}~\bibnamefont {Yin}}, \bibinfo {author}
  {\bibfnamefont {Y.}~\bibnamefont {Wu}}, \bibinfo {author} {\bibfnamefont
  {G.}~\bibnamefont {Li}}, \bibinfo {author} {\bibfnamefont {H.}~\bibnamefont
  {Lei}}, \bibinfo {author} {\bibfnamefont {J.}~\bibnamefont {Li}}, \bibinfo
  {author} {\bibfnamefont {Y.}~\bibnamefont {Chen}}, \bibinfo {author}
  {\bibfnamefont {S.}~\bibnamefont {Yan}},\ and\ \bibinfo {author}
  {\bibfnamefont {Y.}~\bibnamefont {Qi}},\ }\bibfield  {title} {\bibinfo
  {title} {{Charge Density Wave Orders and Enhanced Superconductivity under
  Pressure in the Kagome Metal CsV$_3$Sb$_5$}},\ }\href
  {https://doi.org/10.1002/adma.202102813} {\bibfield  {journal} {\bibinfo
  {journal} {Advanced Materials}\ }\textbf {\bibinfo {volume} {33}},\ \bibinfo
  {pages} {2102813} (\bibinfo {year} {2021}{\natexlab{b}})}\BibitemShut
  {NoStop}%
\bibitem [{\citenamefont {Shumiya}\ \emph {et~al.}(2021)\citenamefont
  {Shumiya}, \citenamefont {Hossain}, \citenamefont {Yin}, \citenamefont
  {Jiang}, \citenamefont {Ortiz}, \citenamefont {Liu}, \citenamefont {Shi},
  \citenamefont {Yin}, \citenamefont {Lei}, \citenamefont {Zhang},
  \citenamefont {Chang}, \citenamefont {Zhang}, \citenamefont {Cochran},
  \citenamefont {Multer}, \citenamefont {Litskevich}, \citenamefont {Cheng},
  \citenamefont {Yang}, \citenamefont {Guguchia}, \citenamefont {Wilson},\ and\
  \citenamefont {Hasan}}]{osti_1821432}%
  \BibitemOpen
  \bibfield  {author} {\bibinfo {author} {\bibfnamefont {N.}~\bibnamefont
  {Shumiya}}, \bibinfo {author} {\bibfnamefont {M.~S.}\ \bibnamefont
  {Hossain}}, \bibinfo {author} {\bibfnamefont {J.-X.}\ \bibnamefont {Yin}},
  \bibinfo {author} {\bibfnamefont {Y.-X.}\ \bibnamefont {Jiang}}, \bibinfo
  {author} {\bibfnamefont {B.~R.}\ \bibnamefont {Ortiz}}, \bibinfo {author}
  {\bibfnamefont {H.}~\bibnamefont {Liu}}, \bibinfo {author} {\bibfnamefont
  {Y.}~\bibnamefont {Shi}}, \bibinfo {author} {\bibfnamefont {Q.}~\bibnamefont
  {Yin}}, \bibinfo {author} {\bibfnamefont {H.}~\bibnamefont {Lei}}, \bibinfo
  {author} {\bibfnamefont {S.~S.}\ \bibnamefont {Zhang}}, \bibinfo {author}
  {\bibfnamefont {G.}~\bibnamefont {Chang}}, \bibinfo {author} {\bibfnamefont
  {Q.}~\bibnamefont {Zhang}}, \bibinfo {author} {\bibfnamefont {T.~A.}\
  \bibnamefont {Cochran}}, \bibinfo {author} {\bibfnamefont {D.}~\bibnamefont
  {Multer}}, \bibinfo {author} {\bibfnamefont {M.}~\bibnamefont {Litskevich}},
  \bibinfo {author} {\bibfnamefont {Z.-J.}\ \bibnamefont {Cheng}}, \bibinfo
  {author} {\bibfnamefont {X.~P.}\ \bibnamefont {Yang}}, \bibinfo {author}
  {\bibfnamefont {Z.}~\bibnamefont {Guguchia}}, \bibinfo {author}
  {\bibfnamefont {S.~D.}\ \bibnamefont {Wilson}},\ and\ \bibinfo {author}
  {\bibfnamefont {M.~Z.}\ \bibnamefont {Hasan}},\ }\bibfield  {title} {\bibinfo
  {title} {{Intrinsic nature of chiral charge order in the kagome
  superconductor RbV$_3$Sb$_5$}},\ }\href
  {https://doi.org/10.1103/physrevb.104.035131} {\bibfield  {journal} {\bibinfo
   {journal} {Physical Review B}\ }\textbf {\bibinfo {volume} {104}},\ \bibinfo
  {pages} {035131} (\bibinfo {year} {2021})}\BibitemShut {NoStop}%
\bibitem [{\citenamefont {Wang}\ \emph
  {et~al.}(2021{\natexlab{c}})\citenamefont {Wang}, \citenamefont {Jiang},
  \citenamefont {Yin}, \citenamefont {Li}, \citenamefont {Wang}, \citenamefont
  {Huang}, \citenamefont {Shao}, \citenamefont {Liu}, \citenamefont {Zhu},
  \citenamefont {Shumiya}, \citenamefont {Hossain}, \citenamefont {Liu},
  \citenamefont {Shi}, \citenamefont {Duan}, \citenamefont {Li}, \citenamefont
  {Chang}, \citenamefont {Dai}, \citenamefont {Ye}, \citenamefont {Xu},
  \citenamefont {Wang}, \citenamefont {Zheng}, \citenamefont {Jia},
  \citenamefont {Hasan},\ and\ \citenamefont {Yao}}]{PhysRevB.104.075148}%
  \BibitemOpen
  \bibfield  {author} {\bibinfo {author} {\bibfnamefont {Z.}~\bibnamefont
  {Wang}}, \bibinfo {author} {\bibfnamefont {Y.-X.}\ \bibnamefont {Jiang}},
  \bibinfo {author} {\bibfnamefont {J.-X.}\ \bibnamefont {Yin}}, \bibinfo
  {author} {\bibfnamefont {Y.}~\bibnamefont {Li}}, \bibinfo {author}
  {\bibfnamefont {G.-Y.}\ \bibnamefont {Wang}}, \bibinfo {author}
  {\bibfnamefont {H.-L.}\ \bibnamefont {Huang}}, \bibinfo {author}
  {\bibfnamefont {S.}~\bibnamefont {Shao}}, \bibinfo {author} {\bibfnamefont
  {J.}~\bibnamefont {Liu}}, \bibinfo {author} {\bibfnamefont {P.}~\bibnamefont
  {Zhu}}, \bibinfo {author} {\bibfnamefont {N.}~\bibnamefont {Shumiya}},
  \bibinfo {author} {\bibfnamefont {M.~S.}\ \bibnamefont {Hossain}}, \bibinfo
  {author} {\bibfnamefont {H.}~\bibnamefont {Liu}}, \bibinfo {author}
  {\bibfnamefont {Y.}~\bibnamefont {Shi}}, \bibinfo {author} {\bibfnamefont
  {J.}~\bibnamefont {Duan}}, \bibinfo {author} {\bibfnamefont {X.}~\bibnamefont
  {Li}}, \bibinfo {author} {\bibfnamefont {G.}~\bibnamefont {Chang}}, \bibinfo
  {author} {\bibfnamefont {P.}~\bibnamefont {Dai}}, \bibinfo {author}
  {\bibfnamefont {Z.}~\bibnamefont {Ye}}, \bibinfo {author} {\bibfnamefont
  {G.}~\bibnamefont {Xu}}, \bibinfo {author} {\bibfnamefont {Y.}~\bibnamefont
  {Wang}}, \bibinfo {author} {\bibfnamefont {H.}~\bibnamefont {Zheng}},
  \bibinfo {author} {\bibfnamefont {J.}~\bibnamefont {Jia}}, \bibinfo {author}
  {\bibfnamefont {M.~Z.}\ \bibnamefont {Hasan}},\ and\ \bibinfo {author}
  {\bibfnamefont {Y.}~\bibnamefont {Yao}},\ }\bibfield  {title} {\bibinfo
  {title} {{Electronic nature of chiral charge order in the kagome
  superconductor $\mathrm{Cs}{\mathrm{V}}_{3}{\mathrm{Sb}}_{5}$}},\ }\href
  {https://doi.org/10.1103/PhysRevB.104.075148} {\bibfield  {journal} {\bibinfo
   {journal} {Phys. Rev. B}\ }\textbf {\bibinfo {volume} {104}},\ \bibinfo
  {pages} {075148} (\bibinfo {year} {2021}{\natexlab{c}})}\BibitemShut
  {NoStop}%
\bibitem [{\citenamefont {Yu}\ \emph {et~al.}(2021{\natexlab{b}})\citenamefont
  {Yu}, \citenamefont {Wu}, \citenamefont {Wang}, \citenamefont {Lei},
  \citenamefont {Zhuo}, \citenamefont {Ying},\ and\ \citenamefont
  {Chen}}]{PhysRevB.104.L041103}%
  \BibitemOpen
  \bibfield  {author} {\bibinfo {author} {\bibfnamefont {F.~H.}\ \bibnamefont
  {Yu}}, \bibinfo {author} {\bibfnamefont {T.}~\bibnamefont {Wu}}, \bibinfo
  {author} {\bibfnamefont {Z.~Y.}\ \bibnamefont {Wang}}, \bibinfo {author}
  {\bibfnamefont {B.}~\bibnamefont {Lei}}, \bibinfo {author} {\bibfnamefont
  {W.~Z.}\ \bibnamefont {Zhuo}}, \bibinfo {author} {\bibfnamefont {J.~J.}\
  \bibnamefont {Ying}},\ and\ \bibinfo {author} {\bibfnamefont {X.~H.}\
  \bibnamefont {Chen}},\ }\bibfield  {title} {\bibinfo {title} {Concurrence of
  anomalous hall effect and charge density wave in a superconducting
  topological kagome metal},\ }\href
  {https://doi.org/10.1103/PhysRevB.104.L041103} {\bibfield  {journal}
  {\bibinfo  {journal} {Phys. Rev. B}\ }\textbf {\bibinfo {volume} {104}},\
  \bibinfo {pages} {L041103} (\bibinfo {year}
  {2021}{\natexlab{b}})}\BibitemShut {NoStop}%
\bibitem [{\citenamefont {Hu}\ \emph {et~al.}(2022{\natexlab{a}})\citenamefont
  {Hu}, \citenamefont {Wu}, \citenamefont {Ortiz}, \citenamefont {Han},
  \citenamefont {Plumb}, \citenamefont {Wilson}, \citenamefont {Schnyder},\
  and\ \citenamefont {Shi}}]{hu2022coexistence}%
  \BibitemOpen
  \bibfield  {author} {\bibinfo {author} {\bibfnamefont {Y.}~\bibnamefont
  {Hu}}, \bibinfo {author} {\bibfnamefont {X.}~\bibnamefont {Wu}}, \bibinfo
  {author} {\bibfnamefont {B.~R.}\ \bibnamefont {Ortiz}}, \bibinfo {author}
  {\bibfnamefont {X.}~\bibnamefont {Han}}, \bibinfo {author} {\bibfnamefont
  {N.~C.}\ \bibnamefont {Plumb}}, \bibinfo {author} {\bibfnamefont {S.~D.}\
  \bibnamefont {Wilson}}, \bibinfo {author} {\bibfnamefont {A.~P.}\
  \bibnamefont {Schnyder}},\ and\ \bibinfo {author} {\bibfnamefont
  {M.}~\bibnamefont {Shi}},\ }\bibfield  {title} {\bibinfo {title}
  {{Coexistence of Tri-Hexagonal and Star-of-David Pattern in the Charge
  Density Wave of the Kagome Superconductor AV$_3$Sb$_5$}},\ }\href
  {https://arxiv.org/abs/2201.06477} {\bibfield  {journal} {\bibinfo  {journal}
  {arXiv preprint}\ }\textbf {\bibinfo {volume} {2201}},\ \bibinfo {pages}
  {06477} (\bibinfo {year} {2022}{\natexlab{a}})}\BibitemShut {NoStop}%
\bibitem [{\citenamefont {Li}\ \emph {et~al.}(2022)\citenamefont {Li},
  \citenamefont {Zhao}, \citenamefont {Ortiz}, \citenamefont {Oey},
  \citenamefont {Wang}, \citenamefont {Wilson},\ and\ \citenamefont
  {Zeljkovic}}]{li2022emergence}%
  \BibitemOpen
  \bibfield  {author} {\bibinfo {author} {\bibfnamefont {H.}~\bibnamefont
  {Li}}, \bibinfo {author} {\bibfnamefont {H.}~\bibnamefont {Zhao}}, \bibinfo
  {author} {\bibfnamefont {B.}~\bibnamefont {Ortiz}}, \bibinfo {author}
  {\bibfnamefont {Y.}~\bibnamefont {Oey}}, \bibinfo {author} {\bibfnamefont
  {Z.}~\bibnamefont {Wang}}, \bibinfo {author} {\bibfnamefont {S.~D.}\
  \bibnamefont {Wilson}},\ and\ \bibinfo {author} {\bibfnamefont
  {I.}~\bibnamefont {Zeljkovic}},\ }\bibfield  {title} {\bibinfo {title}
  {{Emergence of unidirectional coherent quasiparticles from high-temperature
  rotational symmetry broken phase of AV$_3$Sb$_5$ Kagome superconductors}},\
  }\href {https://arxiv.org/abs/2203.15057} {\bibfield  {journal} {\bibinfo
  {journal} {arXiv preprint}\ }\textbf {\bibinfo {volume} {2203}},\ \bibinfo
  {pages} {15057} (\bibinfo {year} {2022})}\BibitemShut {NoStop}%
\bibitem [{\citenamefont {Liang}\ \emph
  {et~al.}(2021{\natexlab{b}})\citenamefont {Liang}, \citenamefont {Hou},
  \citenamefont {Zhang}, \citenamefont {Ma}, \citenamefont {Wu}, \citenamefont
  {Zhang}, \citenamefont {Yu}, \citenamefont {Ying}, \citenamefont {Jiang},
  \citenamefont {Shan}, \citenamefont {Wang},\ and\ \citenamefont
  {Chen}}]{PhysRevX.11.031026}%
  \BibitemOpen
  \bibfield  {author} {\bibinfo {author} {\bibfnamefont {Z.}~\bibnamefont
  {Liang}}, \bibinfo {author} {\bibfnamefont {X.}~\bibnamefont {Hou}}, \bibinfo
  {author} {\bibfnamefont {F.}~\bibnamefont {Zhang}}, \bibinfo {author}
  {\bibfnamefont {W.}~\bibnamefont {Ma}}, \bibinfo {author} {\bibfnamefont
  {P.}~\bibnamefont {Wu}}, \bibinfo {author} {\bibfnamefont {Z.}~\bibnamefont
  {Zhang}}, \bibinfo {author} {\bibfnamefont {F.}~\bibnamefont {Yu}}, \bibinfo
  {author} {\bibfnamefont {J.-J.}\ \bibnamefont {Ying}}, \bibinfo {author}
  {\bibfnamefont {K.}~\bibnamefont {Jiang}}, \bibinfo {author} {\bibfnamefont
  {L.}~\bibnamefont {Shan}}, \bibinfo {author} {\bibfnamefont {Z.}~\bibnamefont
  {Wang}},\ and\ \bibinfo {author} {\bibfnamefont {X.-H.}\ \bibnamefont
  {Chen}},\ }\bibfield  {title} {\bibinfo {title} {{Three-Dimensional Charge
  Density Wave and Surface-Dependent Vortex-Core States in a Kagome
  Superconductor ${\mathrm{CsV}}_{3}{\mathrm{Sb}}_{5}$}},\ }\href
  {https://doi.org/10.1103/PhysRevX.11.031026} {\bibfield  {journal} {\bibinfo
  {journal} {Phys. Rev. X}\ }\textbf {\bibinfo {volume} {11}},\ \bibinfo
  {pages} {031026} (\bibinfo {year} {2021}{\natexlab{b}})}\BibitemShut
  {NoStop}%
\bibitem [{\citenamefont {Jiang}\ \emph {et~al.}(2021)\citenamefont {Jiang},
  \citenamefont {Yin}, \citenamefont {Denner}, \citenamefont {Shumiya},
  \citenamefont {Ortiz}, \citenamefont {Xu}, \citenamefont {Guguchia},
  \citenamefont {He}, \citenamefont {Hossain}, \citenamefont {Liu} \emph
  {et~al.}}]{jiang2021unconventional}%
  \BibitemOpen
  \bibfield  {author} {\bibinfo {author} {\bibfnamefont {Y.-X.}\ \bibnamefont
  {Jiang}}, \bibinfo {author} {\bibfnamefont {J.-X.}\ \bibnamefont {Yin}},
  \bibinfo {author} {\bibfnamefont {M.~M.}\ \bibnamefont {Denner}}, \bibinfo
  {author} {\bibfnamefont {N.}~\bibnamefont {Shumiya}}, \bibinfo {author}
  {\bibfnamefont {B.~R.}\ \bibnamefont {Ortiz}}, \bibinfo {author}
  {\bibfnamefont {G.}~\bibnamefont {Xu}}, \bibinfo {author} {\bibfnamefont
  {Z.}~\bibnamefont {Guguchia}}, \bibinfo {author} {\bibfnamefont
  {J.}~\bibnamefont {He}}, \bibinfo {author} {\bibfnamefont {M.~S.}\
  \bibnamefont {Hossain}}, \bibinfo {author} {\bibfnamefont {X.}~\bibnamefont
  {Liu}}, \emph {et~al.},\ }\bibfield  {title} {\bibinfo {title}
  {{Unconventional chiral charge order in kagome superconductor
  KV$_3$Sb$_5$}},\ }\href {https://www.nature.com/articles/s41563-021-01034-y}
  {\bibfield  {journal} {\bibinfo  {journal} {Nature Materials}\ }\textbf
  {\bibinfo {volume} {20}},\ \bibinfo {pages} {1353} (\bibinfo {year}
  {2021})}\BibitemShut {NoStop}%
\bibitem [{\citenamefont {Hu}\ \emph {et~al.}(2022{\natexlab{b}})\citenamefont
  {Hu}, \citenamefont {Ye}, \citenamefont {Huang}, \citenamefont {Han},
  \citenamefont {Zhao}, \citenamefont {Yang}, \citenamefont {Chen},\ and\
  \citenamefont {Gao}}]{10.1088/1674-1056/ac5888}%
  \BibitemOpen
  \bibfield  {author} {\bibinfo {author} {\bibfnamefont {B.}~\bibnamefont
  {Hu}}, \bibinfo {author} {\bibfnamefont {Y.}~\bibnamefont {Ye}}, \bibinfo
  {author} {\bibfnamefont {Z.}~\bibnamefont {Huang}}, \bibinfo {author}
  {\bibfnamefont {X.}~\bibnamefont {Han}}, \bibinfo {author} {\bibfnamefont
  {Z.}~\bibnamefont {Zhao}}, \bibinfo {author} {\bibfnamefont {H.}~\bibnamefont
  {Yang}}, \bibinfo {author} {\bibfnamefont {H.}~\bibnamefont {Chen}},\ and\
  \bibinfo {author} {\bibfnamefont {H.-J.}\ \bibnamefont {Gao}},\ }\bibfield
  {title} {\bibinfo {title} {{Robustness of the unidirectional stripe order in
  the Kagome superconductor CsV$_3$Sb$_5$}},\ }\href
  {http://iopscience.iop.org/article/10.1088/1674-1056/ac5888} {\bibfield
  {journal} {\bibinfo  {journal} {Chinese Physics B}\ } (\bibinfo {year}
  {2022}{\natexlab{b}})}\BibitemShut {NoStop}%
\bibitem [{\citenamefont {Chen}\ \emph
  {et~al.}(2021{\natexlab{b}})\citenamefont {Chen}, \citenamefont {Yang},
  \citenamefont {Hu}, \citenamefont {Zhao}, \citenamefont {Yuan}, \citenamefont
  {Xing}, \citenamefont {Qian}, \citenamefont {Huang}, \citenamefont {Li},
  \citenamefont {Ye}, \citenamefont {Ma}, \citenamefont {Ni}, \citenamefont
  {Zhang}, \citenamefont {Yin}, \citenamefont {Gong}, \citenamefont {Tu},
  \citenamefont {Lei}, \citenamefont {Tan}, \citenamefont {Zhou}, \citenamefont
  {Shen}, \citenamefont {Dong}, \citenamefont {Yan}, \citenamefont {Wang},\
  and\ \citenamefont {Gao}}]{Chen_2021}%
  \BibitemOpen
  \bibfield  {author} {\bibinfo {author} {\bibfnamefont {H.}~\bibnamefont
  {Chen}}, \bibinfo {author} {\bibfnamefont {H.}~\bibnamefont {Yang}}, \bibinfo
  {author} {\bibfnamefont {B.}~\bibnamefont {Hu}}, \bibinfo {author}
  {\bibfnamefont {Z.}~\bibnamefont {Zhao}}, \bibinfo {author} {\bibfnamefont
  {J.}~\bibnamefont {Yuan}}, \bibinfo {author} {\bibfnamefont {Y.}~\bibnamefont
  {Xing}}, \bibinfo {author} {\bibfnamefont {G.}~\bibnamefont {Qian}}, \bibinfo
  {author} {\bibfnamefont {Z.}~\bibnamefont {Huang}}, \bibinfo {author}
  {\bibfnamefont {G.}~\bibnamefont {Li}}, \bibinfo {author} {\bibfnamefont
  {Y.}~\bibnamefont {Ye}}, \bibinfo {author} {\bibfnamefont {S.}~\bibnamefont
  {Ma}}, \bibinfo {author} {\bibfnamefont {S.}~\bibnamefont {Ni}}, \bibinfo
  {author} {\bibfnamefont {H.}~\bibnamefont {Zhang}}, \bibinfo {author}
  {\bibfnamefont {Q.}~\bibnamefont {Yin}}, \bibinfo {author} {\bibfnamefont
  {C.}~\bibnamefont {Gong}}, \bibinfo {author} {\bibfnamefont {Z.}~\bibnamefont
  {Tu}}, \bibinfo {author} {\bibfnamefont {H.}~\bibnamefont {Lei}}, \bibinfo
  {author} {\bibfnamefont {H.}~\bibnamefont {Tan}}, \bibinfo {author}
  {\bibfnamefont {S.}~\bibnamefont {Zhou}}, \bibinfo {author} {\bibfnamefont
  {C.}~\bibnamefont {Shen}}, \bibinfo {author} {\bibfnamefont {X.}~\bibnamefont
  {Dong}}, \bibinfo {author} {\bibfnamefont {B.}~\bibnamefont {Yan}}, \bibinfo
  {author} {\bibfnamefont {Z.}~\bibnamefont {Wang}},\ and\ \bibinfo {author}
  {\bibfnamefont {H.-J.}\ \bibnamefont {Gao}},\ }\bibfield  {title} {\bibinfo
  {title} {Roton pair density wave in a strong-coupling kagome
  superconductor},\ }\href {https://doi.org/10.1038/s41586-021-03983-5}
  {\bibfield  {journal} {\bibinfo  {journal} {Nature}\ }\textbf {\bibinfo
  {volume} {599}},\ \bibinfo {pages} {222} (\bibinfo {year}
  {2021}{\natexlab{b}})}\BibitemShut {NoStop}%
\bibitem [{\citenamefont {Tan}\ \emph {et~al.}(2021{\natexlab{b}})\citenamefont
  {Tan}, \citenamefont {Liu}, \citenamefont {Wang},\ and\ \citenamefont
  {Yan}}]{PhysRevLett.127.046401}%
  \BibitemOpen
  \bibfield  {author} {\bibinfo {author} {\bibfnamefont {H.}~\bibnamefont
  {Tan}}, \bibinfo {author} {\bibfnamefont {Y.}~\bibnamefont {Liu}}, \bibinfo
  {author} {\bibfnamefont {Z.}~\bibnamefont {Wang}},\ and\ \bibinfo {author}
  {\bibfnamefont {B.}~\bibnamefont {Yan}},\ }\bibfield  {title} {\bibinfo
  {title} {Charge density waves and electronic properties of superconducting
  kagome metals},\ }\href {https://doi.org/10.1103/PhysRevLett.127.046401}
  {\bibfield  {journal} {\bibinfo  {journal} {Phys. Rev. Lett.}\ }\textbf
  {\bibinfo {volume} {127}},\ \bibinfo {pages} {046401} (\bibinfo {year}
  {2021}{\natexlab{b}})}\BibitemShut {NoStop}%
\bibitem [{\citenamefont {Xie}\ \emph {et~al.}(2022)\citenamefont {Xie},
  \citenamefont {Li}, \citenamefont {Bourges}, \citenamefont {Ivanov},
  \citenamefont {Ye}, \citenamefont {Yin}, \citenamefont {Hasan}, \citenamefont
  {Luo}, \citenamefont {Yao}, \citenamefont {Wang}, \citenamefont {Xu},\ and\
  \citenamefont {Dai}}]{PhysRevB.105.L140501}%
  \BibitemOpen
  \bibfield  {author} {\bibinfo {author} {\bibfnamefont {Y.}~\bibnamefont
  {Xie}}, \bibinfo {author} {\bibfnamefont {Y.}~\bibnamefont {Li}}, \bibinfo
  {author} {\bibfnamefont {P.}~\bibnamefont {Bourges}}, \bibinfo {author}
  {\bibfnamefont {A.}~\bibnamefont {Ivanov}}, \bibinfo {author} {\bibfnamefont
  {Z.}~\bibnamefont {Ye}}, \bibinfo {author} {\bibfnamefont {J.-X.}\
  \bibnamefont {Yin}}, \bibinfo {author} {\bibfnamefont {M.~Z.}\ \bibnamefont
  {Hasan}}, \bibinfo {author} {\bibfnamefont {A.}~\bibnamefont {Luo}}, \bibinfo
  {author} {\bibfnamefont {Y.}~\bibnamefont {Yao}}, \bibinfo {author}
  {\bibfnamefont {Z.}~\bibnamefont {Wang}}, \bibinfo {author} {\bibfnamefont
  {G.}~\bibnamefont {Xu}},\ and\ \bibinfo {author} {\bibfnamefont
  {P.}~\bibnamefont {Dai}},\ }\bibfield  {title} {\bibinfo {title}
  {{Electron-phonon coupling in the charge density wave state of
  ${\mathrm{CsV}}_{3}{\mathrm{Sb}}_{5}$}},\ }\href
  {https://doi.org/10.1103/PhysRevB.105.L140501} {\bibfield  {journal}
  {\bibinfo  {journal} {Phys. Rev. B}\ }\textbf {\bibinfo {volume} {105}},\
  \bibinfo {pages} {L140501} (\bibinfo {year} {2022})}\BibitemShut {NoStop}%
\bibitem [{\citenamefont {Ratcliff}\ \emph {et~al.}(2021)\citenamefont
  {Ratcliff}, \citenamefont {Hallett}, \citenamefont {Ortiz}, \citenamefont
  {Wilson},\ and\ \citenamefont {Harter}}]{PhysRevMaterials.5.L111801}%
  \BibitemOpen
  \bibfield  {author} {\bibinfo {author} {\bibfnamefont {N.}~\bibnamefont
  {Ratcliff}}, \bibinfo {author} {\bibfnamefont {L.}~\bibnamefont {Hallett}},
  \bibinfo {author} {\bibfnamefont {B.~R.}\ \bibnamefont {Ortiz}}, \bibinfo
  {author} {\bibfnamefont {S.~D.}\ \bibnamefont {Wilson}},\ and\ \bibinfo
  {author} {\bibfnamefont {J.~W.}\ \bibnamefont {Harter}},\ }\bibfield  {title}
  {\bibinfo {title} {{Coherent phonon spectroscopy and interlayer modulation of
  charge density wave order in the kagome metal
  ${\mathrm{CsV}}_{3}{\mathrm{Sb}}_{5}$}},\ }\href
  {https://doi.org/10.1103/PhysRevMaterials.5.L111801} {\bibfield  {journal}
  {\bibinfo  {journal} {Phys. Rev. Materials}\ }\textbf {\bibinfo {volume}
  {5}},\ \bibinfo {pages} {L111801} (\bibinfo {year} {2021})}\BibitemShut
  {NoStop}%
\bibitem [{\citenamefont {Miao}\ \emph {et~al.}(2021)\citenamefont {Miao},
  \citenamefont {Li}, \citenamefont {Meier}, \citenamefont {Huon},
  \citenamefont {Lee}, \citenamefont {Said}, \citenamefont {Lei}, \citenamefont
  {Ortiz}, \citenamefont {Wilson}, \citenamefont {Yin}, \citenamefont {Hasan},
  \citenamefont {Wang}, \citenamefont {Tan},\ and\ \citenamefont
  {Yan}}]{PhysRevB.104.195132}%
  \BibitemOpen
  \bibfield  {author} {\bibinfo {author} {\bibfnamefont {H.}~\bibnamefont
  {Miao}}, \bibinfo {author} {\bibfnamefont {H.~X.}\ \bibnamefont {Li}},
  \bibinfo {author} {\bibfnamefont {W.~R.}\ \bibnamefont {Meier}}, \bibinfo
  {author} {\bibfnamefont {A.}~\bibnamefont {Huon}}, \bibinfo {author}
  {\bibfnamefont {H.~N.}\ \bibnamefont {Lee}}, \bibinfo {author} {\bibfnamefont
  {A.}~\bibnamefont {Said}}, \bibinfo {author} {\bibfnamefont {H.~C.}\
  \bibnamefont {Lei}}, \bibinfo {author} {\bibfnamefont {B.~R.}\ \bibnamefont
  {Ortiz}}, \bibinfo {author} {\bibfnamefont {S.~D.}\ \bibnamefont {Wilson}},
  \bibinfo {author} {\bibfnamefont {J.~X.}\ \bibnamefont {Yin}}, \bibinfo
  {author} {\bibfnamefont {M.~Z.}\ \bibnamefont {Hasan}}, \bibinfo {author}
  {\bibfnamefont {Z.}~\bibnamefont {Wang}}, \bibinfo {author} {\bibfnamefont
  {H.}~\bibnamefont {Tan}},\ and\ \bibinfo {author} {\bibfnamefont
  {B.}~\bibnamefont {Yan}},\ }\bibfield  {title} {\bibinfo {title} {{Geometry
  of the charge density wave in the kagome metal
  $A{\mathrm{V}}_{3}{\mathrm{Sb}}_{5}$}},\ }\href
  {https://doi.org/10.1103/PhysRevB.104.195132} {\bibfield  {journal} {\bibinfo
   {journal} {Phys. Rev. B}\ }\textbf {\bibinfo {volume} {104}},\ \bibinfo
  {pages} {195132} (\bibinfo {year} {2021})}\BibitemShut {NoStop}%
\bibitem [{\citenamefont {Luo}\ \emph {et~al.}(2021)\citenamefont {Luo},
  \citenamefont {Peng}, \citenamefont {Teicher}, \citenamefont {Huai},
  \citenamefont {Hu}, \citenamefont {Ortiz}, \citenamefont {Wei}, \citenamefont
  {Shen}, \citenamefont {Ou}, \citenamefont {Wang} \emph
  {et~al.}}]{luo2021distinct}%
  \BibitemOpen
  \bibfield  {author} {\bibinfo {author} {\bibfnamefont {Y.}~\bibnamefont
  {Luo}}, \bibinfo {author} {\bibfnamefont {S.}~\bibnamefont {Peng}}, \bibinfo
  {author} {\bibfnamefont {S.~M.}\ \bibnamefont {Teicher}}, \bibinfo {author}
  {\bibfnamefont {L.}~\bibnamefont {Huai}}, \bibinfo {author} {\bibfnamefont
  {Y.}~\bibnamefont {Hu}}, \bibinfo {author} {\bibfnamefont {B.~R.}\
  \bibnamefont {Ortiz}}, \bibinfo {author} {\bibfnamefont {Z.}~\bibnamefont
  {Wei}}, \bibinfo {author} {\bibfnamefont {J.}~\bibnamefont {Shen}}, \bibinfo
  {author} {\bibfnamefont {Z.}~\bibnamefont {Ou}}, \bibinfo {author}
  {\bibfnamefont {B.}~\bibnamefont {Wang}}, \emph {et~al.},\ }\bibfield
  {title} {\bibinfo {title} {{Distinct band reconstructions in kagome
  superconductor CsV$_3$Sb$_5$}},\ }\href {https://arxiv.org/abs/2106.01248}
  {\bibfield  {journal} {\bibinfo  {journal} {arXiv preprint}\ ,\ \bibinfo
  {pages} {01248}} (\bibinfo {year} {2021})}\BibitemShut {NoStop}%
\bibitem [{\citenamefont {Ortiz}\ \emph {et~al.}(2021)\citenamefont {Ortiz},
  \citenamefont {Teicher}, \citenamefont {Kautzsch}, \citenamefont {Sarte},
  \citenamefont {Ratcliff}, \citenamefont {Harter}, \citenamefont {Ruff},
  \citenamefont {Seshadri},\ and\ \citenamefont {Wilson}}]{PhysRevX.11.041030}%
  \BibitemOpen
  \bibfield  {author} {\bibinfo {author} {\bibfnamefont {B.~R.}\ \bibnamefont
  {Ortiz}}, \bibinfo {author} {\bibfnamefont {S.~M.~L.}\ \bibnamefont
  {Teicher}}, \bibinfo {author} {\bibfnamefont {L.}~\bibnamefont {Kautzsch}},
  \bibinfo {author} {\bibfnamefont {P.~M.}\ \bibnamefont {Sarte}}, \bibinfo
  {author} {\bibfnamefont {N.}~\bibnamefont {Ratcliff}}, \bibinfo {author}
  {\bibfnamefont {J.}~\bibnamefont {Harter}}, \bibinfo {author} {\bibfnamefont
  {J.~P.~C.}\ \bibnamefont {Ruff}}, \bibinfo {author} {\bibfnamefont
  {R.}~\bibnamefont {Seshadri}},\ and\ \bibinfo {author} {\bibfnamefont
  {S.~D.}\ \bibnamefont {Wilson}},\ }\bibfield  {title} {\bibinfo {title}
  {{Fermi Surface Mapping and the Nature of Charge-Density-Wave Order in the
  Kagome Superconductor ${\mathrm{CsV}}_{3}{\mathrm{Sb}}_{5}$}},\ }\href
  {https://doi.org/10.1103/PhysRevX.11.041030} {\bibfield  {journal} {\bibinfo
  {journal} {Phys. Rev. X}\ }\textbf {\bibinfo {volume} {11}},\ \bibinfo
  {pages} {041030} (\bibinfo {year} {2021})}\BibitemShut {NoStop}%
\bibitem [{\citenamefont {Zhang}\ \emph {et~al.}(2022)\citenamefont {Zhang},
  \citenamefont {Wang}, \citenamefont {Tsang}, \citenamefont {Liu},
  \citenamefont {Xie}, \citenamefont {Yu}, \citenamefont {Lai},\ and\
  \citenamefont {Goh}}]{https://doi.org/10.48550/arxiv.2202.08570}%
  \BibitemOpen
  \bibfield  {author} {\bibinfo {author} {\bibfnamefont {W.}~\bibnamefont
  {Zhang}}, \bibinfo {author} {\bibfnamefont {L.}~\bibnamefont {Wang}},
  \bibinfo {author} {\bibfnamefont {C.~W.}\ \bibnamefont {Tsang}}, \bibinfo
  {author} {\bibfnamefont {X.}~\bibnamefont {Liu}}, \bibinfo {author}
  {\bibfnamefont {J.}~\bibnamefont {Xie}}, \bibinfo {author} {\bibfnamefont
  {W.~C.}\ \bibnamefont {Yu}}, \bibinfo {author} {\bibfnamefont {K.~T.}\
  \bibnamefont {Lai}},\ and\ \bibinfo {author} {\bibfnamefont {S.~K.}\
  \bibnamefont {Goh}},\ }\bibfield  {title} {\bibinfo {title} {{Emergence of
  large quantum oscillation frequencies in thin flakes of a kagome
  superconductor CsV$_{3}$Sb$_{5}$}},\ }\href
  {https://doi.org/10.48550/ARXIV.2202.08570} {\bibfield  {journal} {\bibinfo
  {journal} {arXiV preprint}\ ,\ \bibinfo {pages} {08570}} (\bibinfo {year}
  {2022})}\BibitemShut {NoStop}%
\bibitem [{\citenamefont {Shrestha}\ \emph {et~al.}(2022)\citenamefont
  {Shrestha}, \citenamefont {Chapai}, \citenamefont {Pokharel}, \citenamefont
  {Miertschin}, \citenamefont {Nguyen}, \citenamefont {Zhou}, \citenamefont
  {Chung}, \citenamefont {Kanatzidis}, \citenamefont {Mitchell}, \citenamefont
  {Welp} \emph {et~al.}}]{shrestha2022nontrivial}%
  \BibitemOpen
  \bibfield  {author} {\bibinfo {author} {\bibfnamefont {K.}~\bibnamefont
  {Shrestha}}, \bibinfo {author} {\bibfnamefont {R.}~\bibnamefont {Chapai}},
  \bibinfo {author} {\bibfnamefont {B.~K.}\ \bibnamefont {Pokharel}}, \bibinfo
  {author} {\bibfnamefont {D.}~\bibnamefont {Miertschin}}, \bibinfo {author}
  {\bibfnamefont {T.}~\bibnamefont {Nguyen}}, \bibinfo {author} {\bibfnamefont
  {X.}~\bibnamefont {Zhou}}, \bibinfo {author} {\bibfnamefont {D.}~\bibnamefont
  {Chung}}, \bibinfo {author} {\bibfnamefont {M.}~\bibnamefont {Kanatzidis}},
  \bibinfo {author} {\bibfnamefont {J.}~\bibnamefont {Mitchell}}, \bibinfo
  {author} {\bibfnamefont {U.}~\bibnamefont {Welp}}, \emph {et~al.},\
  }\bibfield  {title} {\bibinfo {title} {{Nontrivial Fermi surface topology of
  the kagome superconductor CsV$_3$Sb$_5$ probed by de Haas-van Alphen
  oscillations}},\ }\href
  {https://journals.aps.org/prb/abstract/10.1103/PhysRevB.105.024508}
  {\bibfield  {journal} {\bibinfo  {journal} {Physical Review B}\ }\textbf
  {\bibinfo {volume} {105}},\ \bibinfo {pages} {024508} (\bibinfo {year}
  {2022})}\BibitemShut {NoStop}%
\bibitem [{\citenamefont {Fu}\ \emph {et~al.}(2021)\citenamefont {Fu},
  \citenamefont {Zhao}, \citenamefont {Chen}, \citenamefont {Yin},
  \citenamefont {Tu}, \citenamefont {Gong}, \citenamefont {Xi}, \citenamefont
  {Zhu}, \citenamefont {Sun}, \citenamefont {Liu},\ and\ \citenamefont
  {Lei}}]{PhysRevLett.127.207002}%
  \BibitemOpen
  \bibfield  {author} {\bibinfo {author} {\bibfnamefont {Y.}~\bibnamefont
  {Fu}}, \bibinfo {author} {\bibfnamefont {N.}~\bibnamefont {Zhao}}, \bibinfo
  {author} {\bibfnamefont {Z.}~\bibnamefont {Chen}}, \bibinfo {author}
  {\bibfnamefont {Q.}~\bibnamefont {Yin}}, \bibinfo {author} {\bibfnamefont
  {Z.}~\bibnamefont {Tu}}, \bibinfo {author} {\bibfnamefont {C.}~\bibnamefont
  {Gong}}, \bibinfo {author} {\bibfnamefont {C.}~\bibnamefont {Xi}}, \bibinfo
  {author} {\bibfnamefont {X.}~\bibnamefont {Zhu}}, \bibinfo {author}
  {\bibfnamefont {Y.}~\bibnamefont {Sun}}, \bibinfo {author} {\bibfnamefont
  {K.}~\bibnamefont {Liu}},\ and\ \bibinfo {author} {\bibfnamefont
  {H.}~\bibnamefont {Lei}},\ }\bibfield  {title} {\bibinfo {title} {{Quantum
  Transport Evidence of Topological Band Structures of Kagome Superconductor
  ${\mathrm{CsV}}_{3}{\mathrm{Sb}}_{5}$}},\ }\href
  {https://doi.org/10.1103/PhysRevLett.127.207002} {\bibfield  {journal}
  {\bibinfo  {journal} {Phys. Rev. Lett.}\ }\textbf {\bibinfo {volume} {127}},\
  \bibinfo {pages} {207002} (\bibinfo {year} {2021})}\BibitemShut {NoStop}%
\bibitem [{\citenamefont {Chen}\ \emph
  {et~al.}(2021{\natexlab{c}})\citenamefont {Chen}, \citenamefont {He},
  \citenamefont {Yao}, \citenamefont {Pan}, \citenamefont {Lin}, \citenamefont
  {Schnelle}, \citenamefont {Sun}, \citenamefont {Gooth}, \citenamefont
  {Taillefer},\ and\ \citenamefont
  {Felser}}]{https://doi.org/10.48550/arxiv.2110.13085}%
  \BibitemOpen
  \bibfield  {author} {\bibinfo {author} {\bibfnamefont {D.}~\bibnamefont
  {Chen}}, \bibinfo {author} {\bibfnamefont {B.}~\bibnamefont {He}}, \bibinfo
  {author} {\bibfnamefont {M.}~\bibnamefont {Yao}}, \bibinfo {author}
  {\bibfnamefont {Y.}~\bibnamefont {Pan}}, \bibinfo {author} {\bibfnamefont
  {H.}~\bibnamefont {Lin}}, \bibinfo {author} {\bibfnamefont {W.}~\bibnamefont
  {Schnelle}}, \bibinfo {author} {\bibfnamefont {Y.}~\bibnamefont {Sun}},
  \bibinfo {author} {\bibfnamefont {J.}~\bibnamefont {Gooth}}, \bibinfo
  {author} {\bibfnamefont {L.}~\bibnamefont {Taillefer}},\ and\ \bibinfo
  {author} {\bibfnamefont {C.}~\bibnamefont {Felser}},\ }\bibfield  {title}
  {\bibinfo {title} {{Anomalous thermoelectric effects and quantum oscillations
  in the kagome metal CsV$_3$Sb$_5$}},\ }\href
  {https://doi.org/10.48550/ARXIV.2110.13085} {\bibfield  {journal} {\bibinfo
  {journal} {arXiV}\ ,\ \bibinfo {pages} {13085}} (\bibinfo {year}
  {2021}{\natexlab{c}})}\BibitemShut {NoStop}%
\bibitem [{\citenamefont {Singleton}\ \emph {et~al.}(2010)\citenamefont
  {Singleton}, \citenamefont {de~la Cruz}, \citenamefont {McDonald},
  \citenamefont {Li}, \citenamefont {Altarawneh}, \citenamefont {Goddard},
  \citenamefont {Franke}, \citenamefont {Rickel}, \citenamefont {Mielke},
  \citenamefont {Yao},\ and\ \citenamefont {Dai}}]{PhysRevLett.104.086403}%
  \BibitemOpen
  \bibfield  {author} {\bibinfo {author} {\bibfnamefont {J.}~\bibnamefont
  {Singleton}}, \bibinfo {author} {\bibfnamefont {C.}~\bibnamefont {de~la
  Cruz}}, \bibinfo {author} {\bibfnamefont {R.~D.}\ \bibnamefont {McDonald}},
  \bibinfo {author} {\bibfnamefont {S.}~\bibnamefont {Li}}, \bibinfo {author}
  {\bibfnamefont {M.}~\bibnamefont {Altarawneh}}, \bibinfo {author}
  {\bibfnamefont {P.}~\bibnamefont {Goddard}}, \bibinfo {author} {\bibfnamefont
  {I.}~\bibnamefont {Franke}}, \bibinfo {author} {\bibfnamefont
  {D.}~\bibnamefont {Rickel}}, \bibinfo {author} {\bibfnamefont {C.~H.}\
  \bibnamefont {Mielke}}, \bibinfo {author} {\bibfnamefont {X.}~\bibnamefont
  {Yao}},\ and\ \bibinfo {author} {\bibfnamefont {P.}~\bibnamefont {Dai}},\
  }\bibfield  {title} {\bibinfo {title} {{Magnetic Quantum Oscillations in
  ${\mathrm{YBa}}_{2}{\mathrm{Cu}}_{3}{\mathbf{O}}_{6.61}$ and
  ${\mathrm{YBa}}_{2}{\mathrm{Cu}}_{3}{\mathbf{O}}_{6.69}$ in Fields of Up to
  85 T: Patching the Hole in the Roof of the Superconducting Dome}},\ }\href
  {https://doi.org/10.1103/PhysRevLett.104.086403} {\bibfield  {journal}
  {\bibinfo  {journal} {Phys. Rev. Lett.}\ }\textbf {\bibinfo {volume} {104}},\
  \bibinfo {pages} {086403} (\bibinfo {year} {2010})}\BibitemShut {NoStop}%
\bibitem [{\citenamefont {Mikitik}\ and\ \citenamefont
  {Sharlai}(1999)}]{PhysRevLett.82.2147}%
  \BibitemOpen
  \bibfield  {author} {\bibinfo {author} {\bibfnamefont {G.~P.}\ \bibnamefont
  {Mikitik}}\ and\ \bibinfo {author} {\bibfnamefont {Y.~V.}\ \bibnamefont
  {Sharlai}},\ }\bibfield  {title} {\bibinfo {title} {{Manifestation of Berry's
  Phase in Metal Physics}},\ }\href
  {https://doi.org/10.1103/PhysRevLett.82.2147} {\bibfield  {journal} {\bibinfo
   {journal} {Phys. Rev. Lett.}\ }\textbf {\bibinfo {volume} {82}},\ \bibinfo
  {pages} {2147} (\bibinfo {year} {1999})}\BibitemShut {NoStop}%
\bibitem [{\citenamefont {Fuchs}\ \emph {et~al.}(2010)\citenamefont {Fuchs},
  \citenamefont {Pi{\'{e}}chon}, \citenamefont {Goerbig},\ and\ \citenamefont
  {Montambaux}}]{Fuchs_2010}%
  \BibitemOpen
  \bibfield  {author} {\bibinfo {author} {\bibfnamefont {J.~N.}\ \bibnamefont
  {Fuchs}}, \bibinfo {author} {\bibfnamefont {F.}~\bibnamefont
  {Pi{\'{e}}chon}}, \bibinfo {author} {\bibfnamefont {M.~O.}\ \bibnamefont
  {Goerbig}},\ and\ \bibinfo {author} {\bibfnamefont {G.}~\bibnamefont
  {Montambaux}},\ }\bibfield  {title} {\bibinfo {title} {Topological berry
  phase and semiclassical quantization of cyclotron orbits for two dimensional
  electrons in coupled band models},\ }\href
  {https://doi.org/10.1140/epjb/e2010-00259-2} {\bibfield  {journal} {\bibinfo
  {journal} {The European Physical Journal B}\ }\textbf {\bibinfo {volume}
  {77}},\ \bibinfo {pages} {351} (\bibinfo {year} {2010})}\BibitemShut
  {NoStop}%
\bibitem [{\citenamefont {Hu}\ \emph {et~al.}(2017)\citenamefont {Hu},
  \citenamefont {Tang}, \citenamefont {Liu}, \citenamefont {Zhu}, \citenamefont
  {Wei},\ and\ \citenamefont {Mao}}]{hu2017nearly}%
  \BibitemOpen
  \bibfield  {author} {\bibinfo {author} {\bibfnamefont {J.}~\bibnamefont
  {Hu}}, \bibinfo {author} {\bibfnamefont {Z.}~\bibnamefont {Tang}}, \bibinfo
  {author} {\bibfnamefont {J.}~\bibnamefont {Liu}}, \bibinfo {author}
  {\bibfnamefont {Y.}~\bibnamefont {Zhu}}, \bibinfo {author} {\bibfnamefont
  {J.}~\bibnamefont {Wei}},\ and\ \bibinfo {author} {\bibfnamefont
  {Z.}~\bibnamefont {Mao}},\ }\bibfield  {title} {\bibinfo {title} {{Nearly
  massless Dirac fermions and strong Zeeman splitting in the nodal-line
  semimetal ZrSiS probed by de Haas--van Alphen quantum oscillations}},\ }\href
  {https://journals.aps.org/prb/abstract/10.1103/PhysRevB.96.045127} {\bibfield
   {journal} {\bibinfo  {journal} {Physical Review B}\ }\textbf {\bibinfo
  {volume} {96}},\ \bibinfo {pages} {045127} (\bibinfo {year}
  {2017})}\BibitemShut {NoStop}%
\bibitem [{\citenamefont {Zhao}\ \emph
  {et~al.}(2021{\natexlab{b}})\citenamefont {Zhao}, \citenamefont {Li},
  \citenamefont {Ortiz}, \citenamefont {Teicher}, \citenamefont {Park},
  \citenamefont {Ye}, \citenamefont {Wang}, \citenamefont {Balents},
  \citenamefont {Wilson},\ and\ \citenamefont {Zeljkovic}}]{Zhao_2021}%
  \BibitemOpen
  \bibfield  {author} {\bibinfo {author} {\bibfnamefont {H.}~\bibnamefont
  {Zhao}}, \bibinfo {author} {\bibfnamefont {H.}~\bibnamefont {Li}}, \bibinfo
  {author} {\bibfnamefont {B.~R.}\ \bibnamefont {Ortiz}}, \bibinfo {author}
  {\bibfnamefont {S.~M.~L.}\ \bibnamefont {Teicher}}, \bibinfo {author}
  {\bibfnamefont {T.}~\bibnamefont {Park}}, \bibinfo {author} {\bibfnamefont
  {M.}~\bibnamefont {Ye}}, \bibinfo {author} {\bibfnamefont {Z.}~\bibnamefont
  {Wang}}, \bibinfo {author} {\bibfnamefont {L.}~\bibnamefont {Balents}},
  \bibinfo {author} {\bibfnamefont {S.~D.}\ \bibnamefont {Wilson}},\ and\
  \bibinfo {author} {\bibfnamefont {I.}~\bibnamefont {Zeljkovic}},\ }\bibfield
  {title} {\bibinfo {title} {Cascade of correlated electron states in the
  kagome superconductor {CsV}3sb5},\ }\href
  {https://doi.org/10.1038/s41586-021-03946-w} {\bibfield  {journal} {\bibinfo
  {journal} {Nature}\ }\textbf {\bibinfo {volume} {599}},\ \bibinfo {pages}
  {216} (\bibinfo {year} {2021}{\natexlab{b}})}\BibitemShut {NoStop}%
\bibitem [{\citenamefont {Christensen}\ \emph {et~al.}(2021)\citenamefont
  {Christensen}, \citenamefont {Birol}, \citenamefont {Andersen},\ and\
  \citenamefont {Fernandes}}]{Christensen_2021}%
  \BibitemOpen
  \bibfield  {author} {\bibinfo {author} {\bibfnamefont {M.~H.}\ \bibnamefont
  {Christensen}}, \bibinfo {author} {\bibfnamefont {T.}~\bibnamefont {Birol}},
  \bibinfo {author} {\bibfnamefont {B.~M.}\ \bibnamefont {Andersen}},\ and\
  \bibinfo {author} {\bibfnamefont {R.~M.}\ \bibnamefont {Fernandes}},\
  }\bibfield  {title} {\bibinfo {title} {{Theory of the charge density wave in
  AV$_3$Sb$_5$ Kagome metals}},\ }\href
  {https://doi.org/10.1103/physrevb.104.214513} {\bibfield  {journal} {\bibinfo
   {journal} {Physical Review B}\ }\textbf {\bibinfo {volume} {104}},\ \bibinfo
  {pages} {214513} (\bibinfo {year} {2021})}\BibitemShut {NoStop}%
\bibitem [{\citenamefont {Wang}\ \emph {et~al.}(2022)\citenamefont {Wang},
  \citenamefont {Liu}, \citenamefont {Jeon},\ and\ \citenamefont
  {Cho}}]{PhysRevB.105.045135}%
  \BibitemOpen
  \bibfield  {author} {\bibinfo {author} {\bibfnamefont {C.}~\bibnamefont
  {Wang}}, \bibinfo {author} {\bibfnamefont {S.}~\bibnamefont {Liu}}, \bibinfo
  {author} {\bibfnamefont {H.}~\bibnamefont {Jeon}},\ and\ \bibinfo {author}
  {\bibfnamefont {J.-H.}\ \bibnamefont {Cho}},\ }\bibfield  {title} {\bibinfo
  {title} {{Origin of charge density wave in the layered kagome metal
  ${\mathrm{CsV}}_{3}{\mathrm{Sb}}_{5}$}},\ }\href
  {https://doi.org/10.1103/PhysRevB.105.045135} {\bibfield  {journal} {\bibinfo
   {journal} {Phys. Rev. B}\ }\textbf {\bibinfo {volume} {105}},\ \bibinfo
  {pages} {045135} (\bibinfo {year} {2022})}\BibitemShut {NoStop}%
\end{thebibliography}%

\end{document}